\documentclass[11pt]{article}
\usepackage{graphicx}
\usepackage{amsmath}
\usepackage{amsfonts}
\usepackage{amssymb}
\usepackage{epsfig}
\usepackage{times}
\usepackage{calc}
\usepackage{version}
\def\tw{\textwidth}
\begin{document}

%%%%%%%%%%%%%%%%%%%%%%%%%%%%%%%%%%%%%%%%%%%%%%%%%%%%%%%%%%%%%%%%%%%%%%%%%%%%%%

\begin{titlepage}

%%%%%%%%%%%%%%%%%%%%%%%%%%%%%%%%%%%%%%%%%%%%%%%%%%%%%%%%%%%%%%%%%%%%%%%%%%%%%%
July 2006 \hfill

\vskip 5.5cm

\centerline{\bf SINGLE INCLUSIVE DISTRIBUTION AND
TWO-PARTICLE CORRELATIONS INSIDE ONE JET}
\medskip
\centerline{\bf  AT ``MODIFIED LEADING LOGARITHMIC APPROXIMATION''
 OF QUANTUM CHROMODYNAMICS}
\medskip
\centerline{\bf  II : STEEPEST DESCENT EVALUATION AT SMALL $\boldsymbol{X}$}

\vskip 1cm

\centerline{Redamy Perez-Ramos
\footnote{E-mail: perez@lpthe.jussieu.fr}
}

\baselineskip=15pt

\smallskip
\centerline{\em Laboratoire de Physique Th\'eorique et Hautes Energies
\footnote{LPTHE, tour 24-25, 5\raise 3pt \hbox{\tiny \`eme} \'etage,
Universit\'e P. et M. Curie, BP 126, 4 place Jussieu,
F-75252 Paris Cedex 05 (France)}}
\centerline{\em Unit\'e Mixte de Recherche UMR 7589}
\centerline{\em Universit\'e Pierre et Marie Curie-Paris6; CNRS;
Universit\'e Denis Diderot-Paris7}

\vskip 2cm

{\bf Abstract}: The MLLA single inclusive distribution inside one
high energy (gluon) jet at small $x$ is estimated by the steepest
descent method.
Its analytical expression is obtained outside the ``limiting spectrum''.
It is then used to evaluate 2-particle correlations at the same level of
generality. The dependence of both observables on the ratio between the
infrared cutoff $Q_0$ and $\Lambda_{QCD}$ is studied.
Fong \& Webber's results for correlations are recovered at
the limits when this ratio goes to $1$ and when one stays close to
the peak of the single inclusive distribution.

\vskip 1 cm


\vfill

%%%%%%%%%%%%%%%%%%%%%%%%%%%%%%%%%%%%%%%%%%%%%%%%%%%%%%%%%%%%%%%%%%%%%%%%%%%%%

\end{titlepage}

%%%%%%%%%%%%%%%%%%%%%%%%%%%%%%%%%%%%%%%%%%%%%%%%%%%%%%%%%%%%%%%%%%%%%%%%%%%%%

\tableofcontents

\newpage
%%%%%%%%%%%%%%%%%%%%%%%%%%%%%%%%%%%%%%%%%%%%%%%%%%%%%%%%%%%%%%%%%%%%%%%%%%%%%

%%%%%%%%%%%%%%%%%%%%%%%%%%%%%%%%%%%%%%%%%%%%%%%%%%%%%%%%%%%%%%%%%%%%%%%%%%%%%
\section{INTRODUCTION}
\label{section:intro}
%%%%%%%%%%%%%%%%%%%%%%%%%%%%%%%%%%%%%%%%%%%%%%%%%%%%%%%%%%%%%%%%%%%%%%%%%%%%%

Exactly solving the MLLA evolution equations
for the quark and gluon
inclusive spectra and for 2-particle correlations inside one jet
provided, at small $x$, in \cite{RPR2}, analytical
expressions  for these observables, which were unfortunately
limited, for technical reasons to the ``limiting spectrum''
$\lambda \equiv \ln(Q_0/\Lambda_{QCD} =0$. The goal of this second work is
to go beyond this limit in an approximate scheme which proves
very economical and powerful: the steepest descent (SD) method.
It offers sizable technical progress in the calculation of both
observables.

First, we perform a SD evaluation of the (quark and) gluon
single inclusive distributions. Their full dependence on $\lambda$ is
given, including the normalization. The well known  shift to smaller
values of $x$ of the maximum of the distribution, as compared with
DLA calculations is checked, as well as its Gaussian shape
around the maximum. Comparison with the results obtained numerically in
\cite{DKT} is done.

As shown in \cite{RPR2}, knowing the logarithmic derivatives of
the inclusive spectra immediately gives access to 2-particle correlations.
This is accordingly our next step. Since, in particular, the former prove
to be infra-red stable in the limit $\lambda\to 0$, the result can be
safely compared with the exact one obtained in \cite{RPR2}. The agreement
turns out to be excellent, and increases with the energy scale of the
process.

Last, we evaluate 2-particle correlations inside one high energy jet
and study their behavior at $Q_0\ne\Lambda_{QCD}$.
That one recovers the results of Fong \& Webber \cite{FW} close to the peak
of the single inclusive distribution and when $\lambda
\to 0$ is  an important test of the validity and efficiency
of the SD method. The  quantitative predictions do not substantially
differ from the ones of \cite{RPR2} for the ``limiting spectrum'', which
stays  the best candidate to reproduce experimental results.

A conclusion summarizes the achievements, limitations and expectations
of \cite{RPR2} and of the present work. It is completed with two technical
appendices.

%%%%%%%%%%%%%%%%%%%%%%%%%%%%%%%%%%%%%%%%%%%%%%%%%%%%%%%%%%%%%%%%%%%%%%%%%%%%%
\section{STEEPEST DESCENT EVALUATION OF THE SINGLE INCLUSIVE DISTRIBUTION}
\label{sec:SD}
%%%%%%%%%%%%%%%%%%%%%%%%%%%%%%%%%%%%%%%%%%%%%%%%%%%%%%%%%%%%%%%%%%%%%%%%%%%%%

We consider  the production of one hadron inside a quark or a 
gluon jet in a hard process. It carries the fraction $x$ of the total energy
$E$ of the jet. $\Theta_0$ is the half opening angle of the jet while
$\Theta$ is the angle corresponding to the first splitting with
 energy fraction $x\ll z\ll1$.

\subsection{Variables and kinematics}
%%%%%%%%%%%%%%%%%%%%%%%%%%%%%%%%%%%%%

The variables
and  kinematics of the process under consideration are the same as in section
3.1 of \cite{RPR2}.

\subsection{Evolution equations for particle spectra at MLLA}
%%%%%%%%%%%%%%%%%%%%%%%%%%%%%%%%%%%%%%%%%%%%%%%%%%%%%%%%%%%%%%

We define like in \cite{RPR2} the logarithmic parton densities 

$$
Q(\ell)\equiv xD_Q(x),\qquad G(\ell)=xD_G(x)
$$

for quark and gluon jets in terms of which
the system of evolution equations for particle
spectra at small $x$ (see eqs.~(42) and (43) of \cite{RPR2}) read

\begin{equation}
Q(\ell,y)= \delta(\ell) + \frac{C_F}{N_c}\!\!\int_0^\ell d\ell'\int_0^y dy'
\gamma_0^2(\ell'+y')\Big(1
-\frac34\delta(\ell'-\ell) \Big) G(\ell',y'),
\label{eq:solq}
\end{equation}

\begin{equation}
G(\ell,y) = \delta(\ell)
+\int_0^{\ell} d\ell'\int_0^{y} dy' \gamma_0^2(\ell'+y')\Big(
 1  -a\delta(\ell'-\ell) \Big) G(\ell',y'),
\label{eq:solg}
\end{equation}
where
\begin{equation}
a = \frac{1}{4N_c}\bigg[\frac{11}{3}N_c + \frac{4}{3}n_f T_R
 \bigg(\!1-\frac{2C_F}{N_c}\!\bigg)\bigg]\stackrel{n_f=3}{=}0.935.
\label{eq:adef}
\end{equation}
The terms $\propto\frac34$ in (\ref{eq:solq}) and $\propto a$ in (\ref{eq:solg}) 
account for hard
corrections to soft gluon multiplication, sub-leading $g\!\to\!q\bar q$ splittings,
strict angular ordering and energy conservation.

\subsection{Evolution equations; steepest descent evaluation}
\label{subsection:SDeval}
%%%%%%%%%%%%%%%%%%%%%%%%%%%%%%%%%%%%%%%%%%%%%%%%%%%%%%%%%%

The exact solution of (\ref{eq:solg}) is demonstrated in \cite{RPR2} to be given
by the Mellin's integral representation

\begin{eqnarray}\label{eq:MLLAalphasrun}
G\left(\ell,y\right) &=& \left(\ell\!+\!y\!+\!\lambda\right)\!\!\iint
\frac{d\omega\, d\nu}{\left(2\pi i\right)^2}e^{\omega\ell+\nu y}
\!\!\int_{0}^{\infty}\frac{ds}{\nu+s}\!\!\left(\!\frac{\omega
\left(\nu+s\right)}{\left(\omega+s\right)\nu}\!\right)^{1/\beta
\left(\omega-\nu\right)}\!\!\left(\!\frac{\nu}{\nu+s}\!\right)^
{a/\beta}\,e^{-\lambda s}\cr
&=&\left(\ell\!+\!y\!+\!\lambda\right)\!\!\iint
\frac{d\omega\, d\nu}{\left(2\pi i\right)^2}e^{\omega\ell+\nu y}
\!\!\int_{0}^{\infty}\frac{ds}{\nu+s}\left(\!\frac{\nu}{\nu+s}\!\right)^
{a/\beta}e^{\sigma(s)},
\end{eqnarray}

where we have exponentiated the kernel (symmetrical in $(\omega,\nu)$)
\begin{equation}\label{eq:sigma}
\sigma(s)=\frac1{\beta(\omega-\nu)}\ln\left(\!\frac{\omega(\nu+s)}
{\nu(\omega+s)}\!\right)-\lambda s.
\end{equation}

(\ref{eq:MLLAalphasrun}) will be estimated by the SD method.
The value $s_0$ of the saddle point, satisfying
$\frac{d\sigma(s)}{ds}\Big\vert_{s=s_0}=0$, reads (see \cite{DLA})

\begin{equation}\label{eq:saddlepoint}
s_0(\omega,\nu)=\frac12\left[\sqrt{\frac4{\beta\lambda}+(\omega-\nu)^2}-(\omega+\nu)
\right].
\end{equation}
One makes a Taylor expansion of $\sigma(s)$ nearby $s_0$:

\begin{equation}\label{eq:sigmas0}
\sigma(s)=\sigma(s_0)+\frac12\sigma''(s_0)(s-s_0)^2+{\cal O}\left((s-s_0)^2\right),
\quad
\sigma''(s_0)=-\beta\lambda^2\sqrt{\frac{4}{\beta\lambda}+(\omega-\nu)^2}<0,
\end{equation}

such that

\begin{eqnarray}\label{eq:sigmavalue}
\int_{0}^{\infty}\frac{ds}{\nu+s}\!\!\left(\!\frac{\omega
\left(\nu+s\right)}{\left(\omega+s\right)\nu}\!\right)^{1/\beta
\left(\omega-\nu\right)}\!\!\left(\!\frac{\nu}{\nu+s}\!\right)^
{a/\beta}\,e^{-\lambda s}
\!\!&\!\!\stackrel{\lambda\gg1}{\approx}\!\!&\!\!
2\sqrt{\frac{\pi}2}\displaystyle{\frac{e^{\sigma(s_0)}}{(\nu+s_0)
\sqrt{\mid\!\sigma''(s_0)\!\mid}}}\left(\!\frac{\nu}{\nu+s_0}\!\right)^{a/\beta}\!\!.
\end{eqnarray}

The condition $\lambda\!\gg\!1$$\Rightarrow$$\alpha_s/\pi\!\ll\!1$ 
\footnote{in (\ref{eq:sigmas0}), $\lambda$ appears to the power $3/2>1$,
which guarantees the fast convergence of the SD as $\lambda$ increases.}
guarantees, in particular, the convergence of the perturbative approach. Substituting (\ref{eq:sigmavalue}) in 
(\ref{eq:MLLAalphasrun}) yields

\begin{equation}\label{eq:SpecDD}
G\left(\ell,y\right)\approx2\sqrt{\frac{\pi}2}(\ell+y+\lambda)
\iint\frac{d\omega\, d\nu}{\left(2\pi i\right)^2}\,
\displaystyle{\frac{e^{\phi\left(\omega,\nu,\ell,y\right)}}{(\nu+s_0)
\sqrt{\mid\!\sigma''(s_0)\!\mid}}}\left(\!\frac{\nu}{\nu+s_0}\!\right)^{a/\beta},
\end{equation}

where the argument of the exponential is

\begin{equation}\label{eq:phiexp}
\phi\left(\omega,\nu,\ell,y\right)=\omega\ell+\nu y+
\frac1{\beta\left(\omega-\nu\right)}
\ln{\frac{\omega\left(\nu+s_0\right)}{\left(\omega+s_0\right)\nu}}-
\lambda s_0.
\end{equation}

Once again, we perform the SD method to evaluate 
(\ref{eq:SpecDD}). The saddle point $(\omega_0,\nu_0)$ satisfies
the equations

\vbox{
\begin{subequations}
\begin{equation}\label{eq:deromega}
\frac{\partial\phi}{\partial\omega}=\ell-\frac1{\beta\left(\omega-\nu\right)^2}
\ln{\frac{\omega\left(\nu+s_0\right)}{\left(\omega+s_0\right)\nu}}+\frac1
{\beta\omega\left(\omega-\nu\right)}-\lambda\frac{\left(\nu+s_0\right)}
{\left(\omega-\nu\right)}=0,
\end{equation}

\begin{equation}\label{eq:dernu}
\frac{\partial\phi}{\partial\nu}=y+\frac1{\beta\left(\omega-\nu\right)^2}
\ln{\frac{\omega\left(\nu+s_0\right)}{\left(\omega+s_0\right)\nu}}-
\frac1{\beta\nu\left(\omega-\nu\right)}+\lambda
\frac{\left(\omega+s_0\right)}{\left(\omega-\nu\right)}=0.
\end{equation}
\end{subequations}
}

Adding and subtracting (\ref{eq:deromega}) and (\ref{eq:dernu}) gives respectively

\begin{subequations}
\begin{equation}
\omega_0\nu_0=\frac1{\beta\left(\ell+y+\lambda\right)},
\end{equation}

\begin{equation}\label{eq:yml}
y-\ell=\frac1{\beta\left(\omega_0-\nu_0\right)}
\left(\frac1{\omega_0}+\frac1{\nu_0}\right)-\frac2
{\beta\left(\omega_0-\nu_0\right)^2}\ln{\frac{\omega_0
\left(\nu_0+s_0\right)}{\left(\omega_0+s_0\right)\nu_0}}-
\lambda\frac{\omega_0+\nu_0+2s_0}{\omega_0-\nu_0};
\end{equation}
\end{subequations}

$(\omega_0,\nu_0)$ also satisfies (from (\ref{eq:saddlepoint}))

\begin{equation}
\left(\omega_0+s_0\right)\left(\nu_0+s_0\right)=\frac1{\beta\lambda}.
\end{equation}

One can substitute the expressions
(\ref{eq:deromega}) and (\ref{eq:dernu}) of  $\ell$ and $y$ 
into (\ref{eq:phiexp}), which yields

\begin{equation}\label{eq:phiexpbis}\
\varphi\equiv\phi(\omega_0,\nu_0,\ell,y)=\frac{2}
{\beta\left(\omega_0-\nu_0\right)}\ln{\frac{\omega_0\left(\nu_0+s_0\right)}
{\left(\omega_0+s_0\right)\nu_0}}.
\end{equation}

Introducing the variables $(\mu,\upsilon)$ \cite{DLA} to parametrize
$(\omega_0,\nu_0)$ through
\begin{equation}\label{eq:muupsilon}
\omega_0\left(\nu_0\right)=\frac1{\sqrt{\beta(\ell\!+\!y\!+\!\lambda)}}
e^{\pm\mu(\ell,y)},\qquad \left(\omega_0+s_0\right)
\left(\nu_0+s_0\right)=\frac1{\sqrt{\beta\lambda}}e^{\pm\upsilon(\ell,y)},
\end{equation}
one rewrites (\ref{eq:phiexpbis}) and (\ref{eq:yml}) respectively
in the form
\begin{equation}\label{eq:phi}
\varphi(\mu,\upsilon)=\frac2{\sqrt{\beta}}\left(\sqrt{\ell+y+\lambda}-\sqrt{\lambda}\right)
\,\frac{\mu-\upsilon}{\sinh\mu-\sinh\upsilon},
\end{equation}

\begin{subequations}
\begin{equation}\label{eq:ratiomunu}
\frac{y-\ell}{y+\ell}=\frac{\left(\sinh 2\mu-2\mu\right)-\left(\sinh 2\upsilon-2\upsilon\right)}{2\left(\sinh^2\mu-\sinh^2\upsilon\right)};
\end{equation}

moreover, since $\omega_0-\nu_0=(\omega_0-s_0)-(\nu_0-s_0)$, $(\mu,\upsilon)$ also
satisfy

\begin{equation}\label{eq:relmunu}
\frac{\sinh\upsilon}{\sqrt{\lambda}}=\frac{\sinh\mu}{\sqrt{\ell+y+\lambda}}.
\end{equation}
\end{subequations}

Performing a Taylor expansion of $\phi(\omega,\nu,\ell,y)$ around $(\omega_0,\nu_0)$,
which needs evaluating $\frac{\partial^2\phi}{\partial\omega^2}$, $\frac{\partial^2\phi}{\partial\nu^2}$ and 
$\frac{\partial^2\phi}{\partial\omega\partial\nu}$ (see appendix \ref{sec:DDDet}),
treating $(Y+\lambda)$ as a large parameter and making use of 
(\ref{eq:muupsilon}) provides the SD result

\begin{eqnarray}
G(\ell,y)\approx{\cal N}(\mu,\upsilon,\lambda)\exp\Big[
\frac2{\sqrt{\beta}}\left(\sqrt{\ell+y+\lambda}-\sqrt{\lambda}\right)
\frac{\mu-\upsilon}{\sinh\mu-\sinh\upsilon}+\upsilon
-\frac{a}{\beta}(\mu-\upsilon)\Big],
\label{eq:Specalphasrunmlla}
\end{eqnarray}

where
$$
{\cal N}(\mu,\upsilon,\lambda)=\frac12(\ell\!+\!y\!+\!\lambda)
\frac{\left(\frac{\beta}{\lambda}\right)^{1/4}}{\sqrt{\pi\cosh\upsilon\,
DetA(\mu,\upsilon)}}\left(\frac{\lambda}{\ell+y+\lambda}\right)^{a/2\beta}
$$

with (see details in appendix \ref{sec:DDDet})

\begin{equation}\label{eq:determinant}
DetA(\mu,\nu)=\beta\,(\ell\!+\!y\!+\!\lambda)^3
\left[\frac{(\mu\!-\!\upsilon)\cosh\mu\cosh\upsilon\!+\!\cosh\mu\sinh\upsilon
\!-\!\sinh\mu\cosh\upsilon}
{\sinh^3\mu\cosh\upsilon}\right].
\end{equation}

\subsubsection{Shape of the spectrum given in eq.~(\ref{eq:Specalphasrunmlla})}
\label{subsub:shape}
%%%%%%%%%%%%%%%%%%%%%%%%%%%%%%%%%%%%%%%%%%%%%%%%%%%%%%%%%%%%%%%%%%

We normalize (\ref{eq:Specalphasrunmlla}) by
the MLLA mean multiplicity inside one jet \cite{EvEq}
$$
\bar{n}(Y)\stackrel{\lambda\gg1}{\approx}\frac12\left(\frac{Y+\lambda}{\lambda}\right)^
{\displaystyle{-\frac12\frac{a}{\beta}+\frac14}}
\exp\left[\frac2{\sqrt{\beta}}\left(\sqrt{Y+\lambda}-\sqrt{\lambda}\right)\right].
$$
The normalized expression for the single inclusive distribution
as a function of $\ell=\ln(1/x)$ is accordingly
 obtained by setting $y=Y-\ell$ in (\ref{eq:Specalphasrunmlla})

\begin{eqnarray}
\frac{G(\ell,Y)}{\bar{n}(Y)}\approx\sqrt{\frac{\beta^{1/2}(Y\!+\!\lambda)^{3/2}}
{\pi\cosh\upsilon DetA(\mu,\upsilon)}}\!\exp\!\Big[
\frac2{\sqrt{\beta}}\left(\sqrt{Y\!+\!\lambda}\!-\!\sqrt{\lambda}\right)
\!\left(\frac{\mu\!-\!\upsilon}{\sinh\mu\!-\!\sinh\upsilon}\!-\!1\right)\!+\!\upsilon
\!-\!\frac{a}{\beta}
(\mu\!-\!\upsilon)\Big].
\label{eq:SpecNormMLLA}
\end{eqnarray}

One can explicitly verify that (\ref{eq:SpecNormMLLA}) preserves the position 
of the maximum \cite{EvEq}\cite{KO}\cite{FW1} at

\begin{equation}\label{eq:ellmaxmlla}
\ell_{max}=\frac{Y}2+\frac12\frac{a}{\beta}
\left(\sqrt{Y+\lambda}-\sqrt{\lambda}\right)>\frac{Y}2,
\end{equation}
as well as the gaussian shape of the distribution around (\ref{eq:ellmaxmlla}) 
(see appendix \ref{subsec:DeTA})
\begin{equation}\label{eq:gaussian}
\frac{G(\ell,Y)}{\bar{n}(Y)}\approx\left(\frac{3}{\pi\sqrt{\beta}
\left[(Y+\lambda)^{3/2}
-\lambda^{3/2}\right]}\right)^{1/2}\!\!\exp
\left(-\frac2{\sqrt{\beta}}\,\frac3{(Y+\lambda)^{3/2}
-\lambda^{3/2}}\frac{\left(\ell-\ell_{max}\right)^2}2\right).
\end{equation}

\begin{figure}[h]
\begin{center}
\epsfig{file=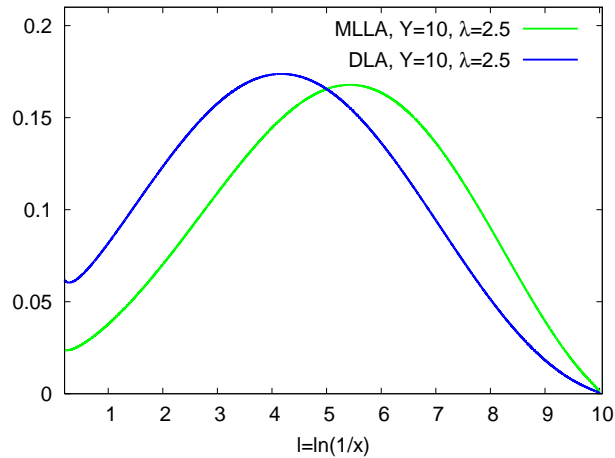, height=6truecm,width=0.5\tw}
\caption{\label{fig:MLLAlambda}SD normalized spectrum: DLA (blue),  MLLA 
 (green);  $Y=10.0$, $\lambda=2.5$.}
\end{center}
\end{figure}

In Fig.\ref{fig:MLLAlambda} we compare for $Y=10$ and $\lambda=2.5$ 
the MLLA curve with DLA (by setting $a=0$ in (\ref{eq:SpecNormMLLA})). 
The general features of the MLLA curve (\ref{eq:SpecNormMLLA}) 
at $\lambda\ne0$ are in good agreement with those of \cite{DKT}.

The shape of the single inclusive spectrum given by (\ref{eq:SpecNormMLLA}) 
can easily be proved to be ``infrared stable'' (it has indeed a final limit
when $\lambda\to0$).

\subsection{Logarithmic derivatives}
%%%%%%%%%%%%%%%%%%%%%%%%%%%%%%%%%%%%

Their calculation is important since they appear in the expressions of
2-particle correlations.

Exponentiating the $(\ell,y)$ dependence of the factor ${\cal N}$ in (\ref{eq:Specalphasrunmlla}), we decompose the whole expression in two pieces
\begin{equation}
\psi=\varphi+\delta\psi,
\label{eq:SP}
\end{equation}

where $\varphi$, given in (\ref{eq:phi}), is the DLA term for the shape of 
the distribution \cite{DLA}, and
\begin{equation}\label{eq:phiprime}
\delta\psi=-\frac12\left(1+\frac{a}{\beta}\right)\ln(\ell+y+\lambda)-\frac{a}{\beta}\mu
+\left(1+\frac{a}{\beta}\right)\upsilon+\frac12\ln[Q(\mu,\upsilon)]
\end{equation}

is the sub-leading contribution (in the sense that its derivative gives
the MLLA correction), where

$$
Q(\mu,\upsilon)\equiv
\frac{\beta(\ell+y+\lambda)^3}{\cosh\upsilon Det\,A(\mu,\upsilon)}=\frac{\sinh^3\mu}
{(\mu\!-\!\upsilon)\cosh\mu\cosh\upsilon\!+\!\cosh\mu\sinh\upsilon\!-\!\sinh\mu
\cosh\upsilon}.
$$

By the definition of the saddle point, the derivatives of
(\ref{eq:phi}) over $\ell$ and $y$ respectively read:

\begin{equation}\label{eq:SPbis}
\varphi_{\ell}=\omega_0=\gamma_0e^{\mu},\qquad \qquad
\varphi_{y}=\nu_0=\gamma_0e^{-\mu}.
\end{equation}

\medskip

We introduce (see appendix \ref{subsec:LK})
\begin{eqnarray}
&&{\cal {L}}(\mu,\upsilon)=-\frac{a}{\beta}+L(\mu,\upsilon),\qquad\quad
L(\mu,\upsilon)=\frac12\frac{\partial}{\partial\mu}
\ln[Q(\mu,\upsilon)],\cr\cr\cr
&&{\cal {K}}(\mu,\upsilon)=1+\frac{a}{\beta}+K(\mu,\upsilon),\qquad\quad
K(\mu,\upsilon)=\frac12\frac{\partial}{\partial\upsilon}
\ln[Q(\mu,\upsilon)]\label{eq:LK}
\end{eqnarray}
and make use of

$$
\frac{\partial\upsilon}{\partial\ell}=\tanh\upsilon
\left(\coth\mu\frac{\partial\mu}{\partial\ell}-
\frac12\beta\gamma_0^2\right),\qquad\quad
\frac{\partial\upsilon}{\partial y}=\tanh\upsilon
\left(\coth\mu\frac{\partial\mu}{\partial y}-
\frac12\beta\gamma_0^2\right),
$$

that follows from (\ref{eq:relmunu}), to write 
$\delta\psi_{\ell},\,\,\delta\psi_y$ in terms of $\frac{\partial\mu}{\partial\ell}$, $\frac{\partial\mu}{\partial y}$

\begin{subequations}
\begin{eqnarray}\label{eq:derpsiprime1}
&&\hskip -0.5cm\delta\psi_{\ell}=-\frac12\left(1+\frac{a}{\beta}
+\tanh\upsilon\,{\cal {K}}(\mu,\upsilon)\right)\beta\gamma_0^2+
\bigg({\cal {L}}(\mu,\upsilon)+\tanh{\upsilon}
\coth{\mu}\,{\cal {K}}(\mu,\upsilon)\bigg)\frac{\partial\mu}{\partial\ell},
\\\notag\\
&&\hskip -0.5cm\delta\psi_{y}=-\frac12\left(1+\frac{a}{\beta}
+\tanh\upsilon\,{\cal {K}}(\mu,\upsilon)\right)\beta\gamma_0^2+
\bigg({\cal {L}}(\mu,\upsilon)+\tanh{\upsilon}
\coth{\mu}\,{\cal {K}}(\mu,\upsilon)\bigg)\frac{\partial\mu}{\partial y}
\label{eq:derpsiprime1bis}.
\end{eqnarray}
\end{subequations}

\begin{figure}[h]
\begin{center}
\epsfig{file=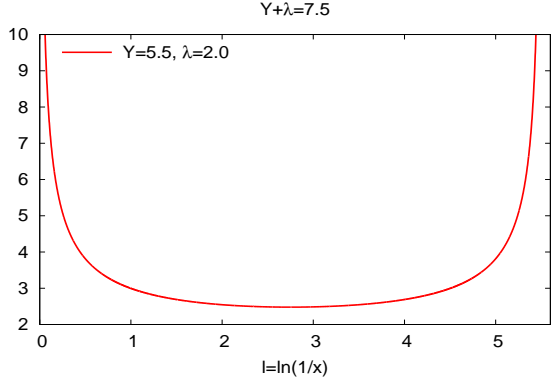, height=5truecm,width=0.45\tw}
\caption{\label{fig:tildeQ} Behavior of $\widetilde Q(\mu,\upsilon)$ as a function
of $\ell=\ln(1/x)$.}
\end{center}
\end{figure}

Using (\ref{eq:ratiomunu}) and (\ref{eq:relmunu}) we obtain

\begin{equation}\label{eq:dermul1}
\frac{\partial\mu}{\partial\ell}=-\frac12\beta\gamma_0^2
\left[1+e^{\mu}\widetilde{Q}(\mu,\upsilon)\right],\qquad\quad
\frac{\partial\mu}{\partial y}=\frac12\beta\gamma_0^2
\left[1+e^{-\mu}\widetilde{Q}(\mu,\upsilon)\right]
\end{equation}

where

\begin{equation}\label{eq:tildeQ}
\widetilde{Q}(\mu,\upsilon)=\frac{\cosh\mu\sinh\mu\cosh\upsilon-
(\mu-\upsilon)\cosh\upsilon-\sinh\upsilon}
{(\mu-\upsilon)\cosh\mu\cosh\upsilon+\cosh\mu\sinh\upsilon-\sinh\mu\cosh\upsilon},
\end{equation}

which we have displayed in Fig.\ref{fig:tildeQ} (useful for correlations).

\begin{figure}[h]
\begin{center}
\epsfig{file=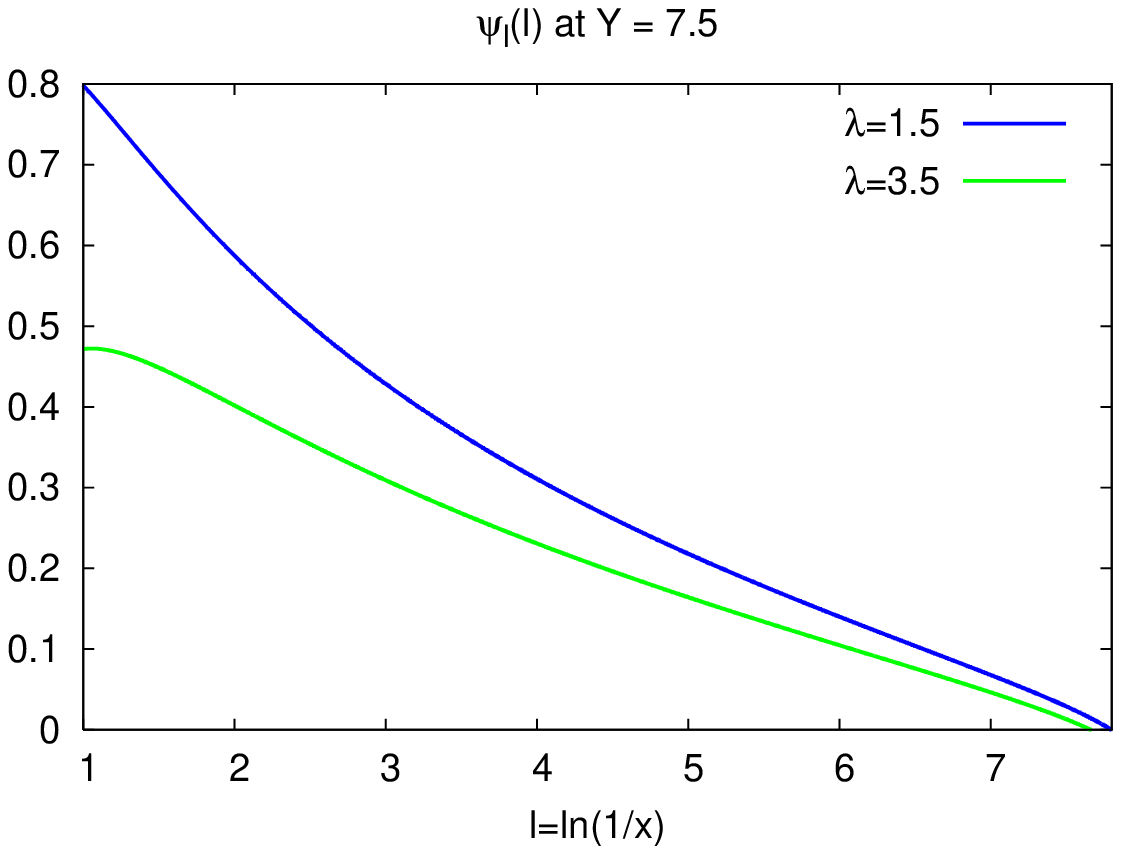, height=5truecm,width=0.48\tw}
\hfill
\epsfig{file=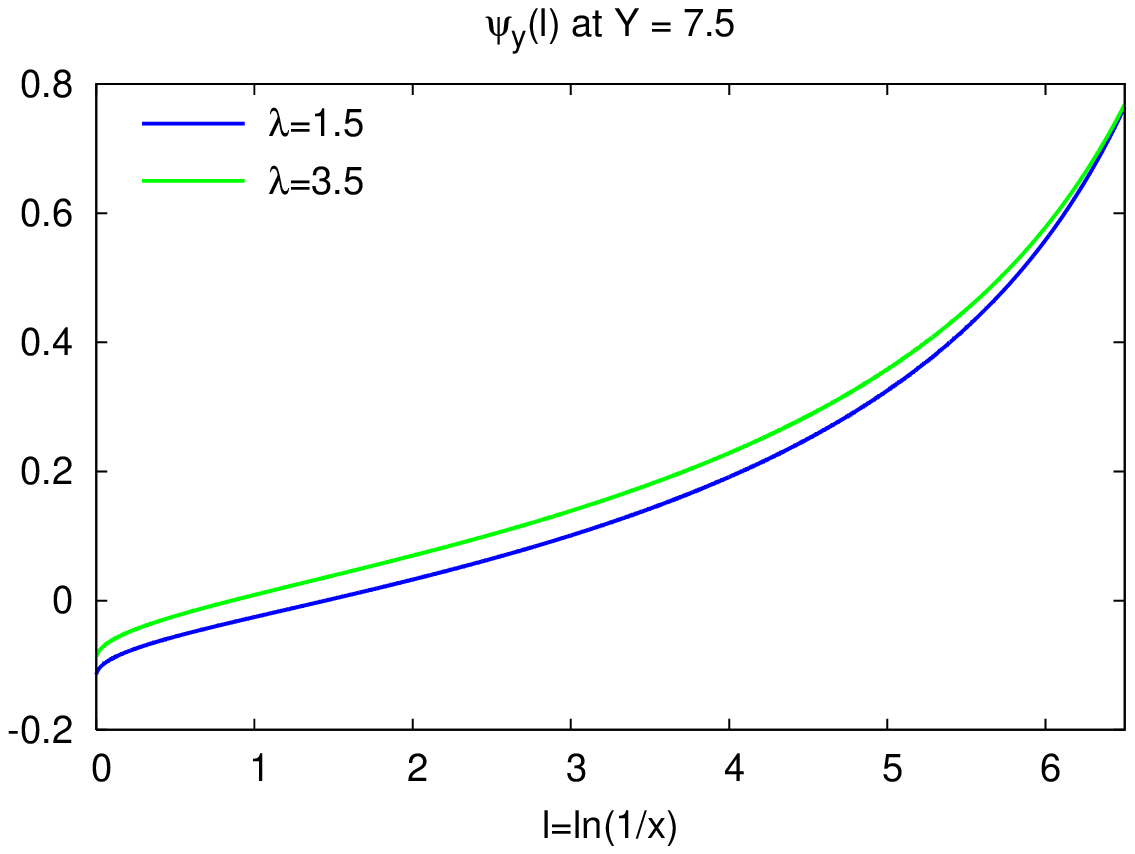, height=5truecm,width=0.48\tw}
\caption{\label{fig:psiellylambda} SD logarithmic derivatives $\psi_\ell$
and $\psi_y$ of the inclusive spectrum at $Y=7.5$,
for $\lambda=1.5$ and $\lambda=3.5$.}
\end{center}
\end{figure}

Inserting (\ref{eq:derpsiprime1bis}) and (\ref{eq:dermul1}) into (\ref{eq:derpsiprime1})\label{eq:derpsiprime2} 
gives the SD logarithmic derivatives of the single inclusive distribution

\begin{subequations}
\begin{eqnarray}
\psi_{\ell}(\mu,\upsilon)\!\!&\!\!=\!\!&\!\!\gamma_0e^{\mu}\!+\!
\frac12a\gamma_0^2\Big[e^{\mu}\widetilde{Q}(\mu,\upsilon)\!-\!\tanh\upsilon
\!-\!\tanh\upsilon\coth\mu\Big(1\!+\!e^{\mu}\widetilde{Q}(\mu,\upsilon)\Big)\Big]\cr\cr
&&\hskip 0.3cm-\frac12\beta\gamma_0^2\Big[1\!+\!\tanh\upsilon
\Big(1\!+\!K(\mu,\upsilon)\Big)+C(\mu,\upsilon)
\Big(1\!+\!e^{\mu}\widetilde{Q}(\mu,\upsilon)\Big)\Big]+{\cal O}(\gamma_0^2)
\label{eq:derpsi'l},
\end{eqnarray}
\begin{eqnarray}
\psi_{y}(\mu,\upsilon)\!\!&\!\!=\!\!&\!\!\gamma_0e^{-\mu}\!-\!\frac12a\gamma_0^2
\Big[2\!+\!e^{-\mu}\widetilde{Q}(\mu,\upsilon)\!+\!\tanh\upsilon
\!-\!\tanh\upsilon\coth\mu\Big(1\!+\!e^{-\mu}\widetilde{Q}(\mu,\upsilon)\Big)\Big]\cr\cr
&&\hskip 0.6cm-\frac12\beta\gamma_0^2\Big[1\!+\!\tanh\upsilon
\Big(1\!+\!K(\mu,\upsilon)\Big)-C(\mu,\upsilon)
\Big(1\!+\!e^{-\mu}\widetilde{Q}(\mu,\upsilon)\Big)\Big]+{\cal O}(\gamma_0^2)
\label{eq:derpsi'y}
\end{eqnarray}
\end{subequations}

where we have introduced ($L$ and $K$ have been written in (\ref{eq:LL})
and (\ref{eq:KK}))

\begin{equation}\label{eq:CC}
C(\mu,\upsilon)=L(\mu,\upsilon)+
\tanh\upsilon\coth\mu\Big(1 + K(\mu,\upsilon)\Big).
\end{equation}
$C$ does not diverge  when $\mu\sim\upsilon\to0$. One has indeed
$$
\lim_{\mu,\upsilon\rightarrow0}\left[L(\mu,\upsilon)+
\tanh\upsilon\coth\mu K(\mu,\upsilon)\right]=\lim_{\mu,\upsilon\rightarrow0}
\frac{2-3\frac{\upsilon^2}{\mu^2}-\frac{\upsilon^3}{\mu^3}}{4\left(1-\frac{\upsilon^3}
{\mu^3}\right)}\mu=0
$$
as well as
\begin{equation*}
\lim_{\mu,\upsilon\rightarrow0}
\tanh\upsilon\coth\mu\left(1+e^{\pm\mu}\widetilde{Q}(\mu,\upsilon)\right)=
\lim_{\mu,\upsilon\rightarrow0}=\frac{3\frac{\upsilon}{\mu}}{1-\frac{\upsilon^3}
{\mu^3}}=\frac{3\sqrt{\frac{\lambda}{Y+\lambda}}}{1-
\left(\frac{\lambda}{Y+\lambda}\right)^{3/2}}.
\end{equation*}
In (\ref{eq:derpsi'l}) and (\ref{eq:derpsi'y}) it is easy to keep trace of
leading and sub-leading contributions.
The first ${\cal O}(\gamma_0)$ term is DLA \cite{DLA}
while the second ($\propto a\to$ ``hard corrections'') and third 
($\propto\beta\to$ ``running coupling effects'') terms
are MLLA corrections (${\cal O}(\gamma_0^2)$), of relative order
${\cal O}(\gamma_0)$ with respect to the leading one. In Fig.\ref{fig:psiellylambda}
we plot (\ref{eq:derpsi'l}) (left) and (\ref{eq:derpsi'y}) (right) for two 
different values of $\lambda$; one observes that $\psi_\ell$ 
($\psi_y$) decreases (increases) when $\lambda$ increases.

For further use in correlations, the  logarithmic derivatives
have the important property that they do not depend on the normalization
but only on the  shape of the single inclusive distribution.

\subsubsection{``Limiting spectrum'': $\boldsymbol{\lambda\to0\ (Q_0
=\Lambda_{QCD})}$}
%%%%%%%%%%%%%%%%%%%%%%%%%%%%%%%%%%%%%%%%%%%%%%%%%%%%%%%%%%%%%%%%%%

\begin{figure}[h]
\begin{center}
\epsfig{file=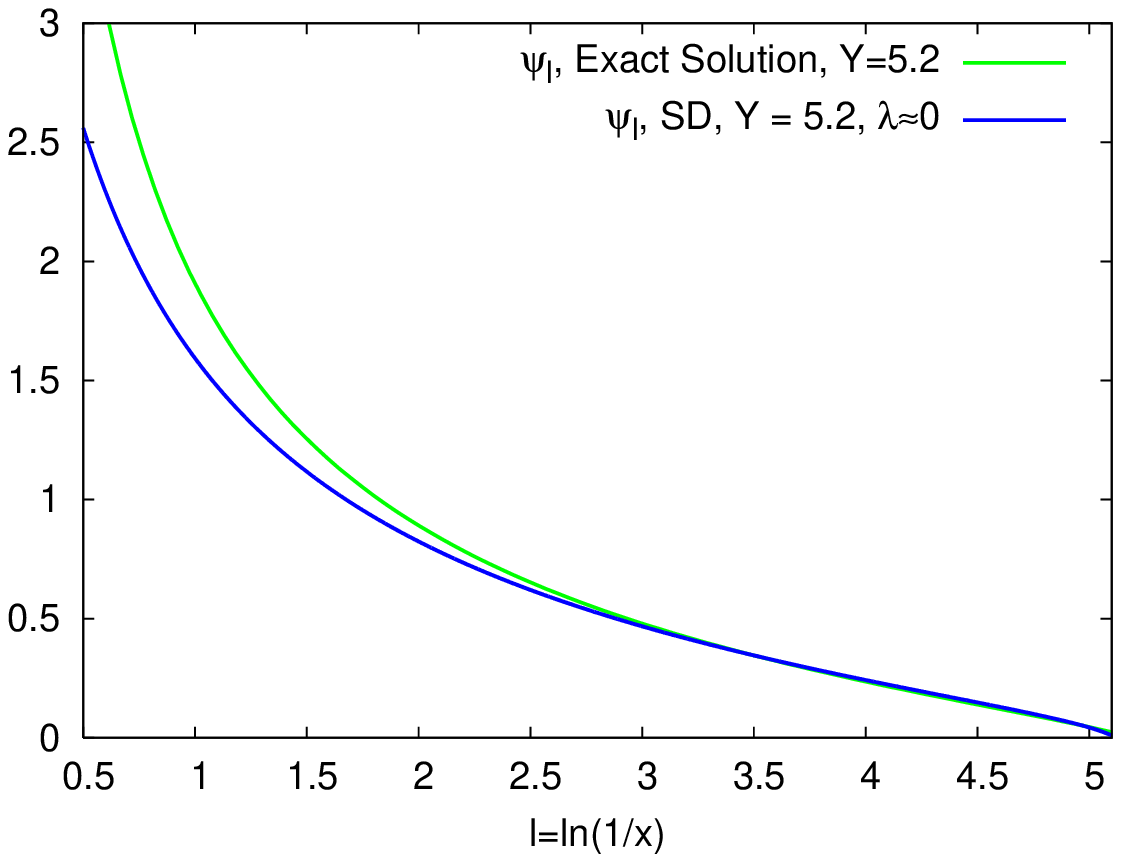, height=5truecm,width=0.48\tw}
\hfill
\epsfig{file=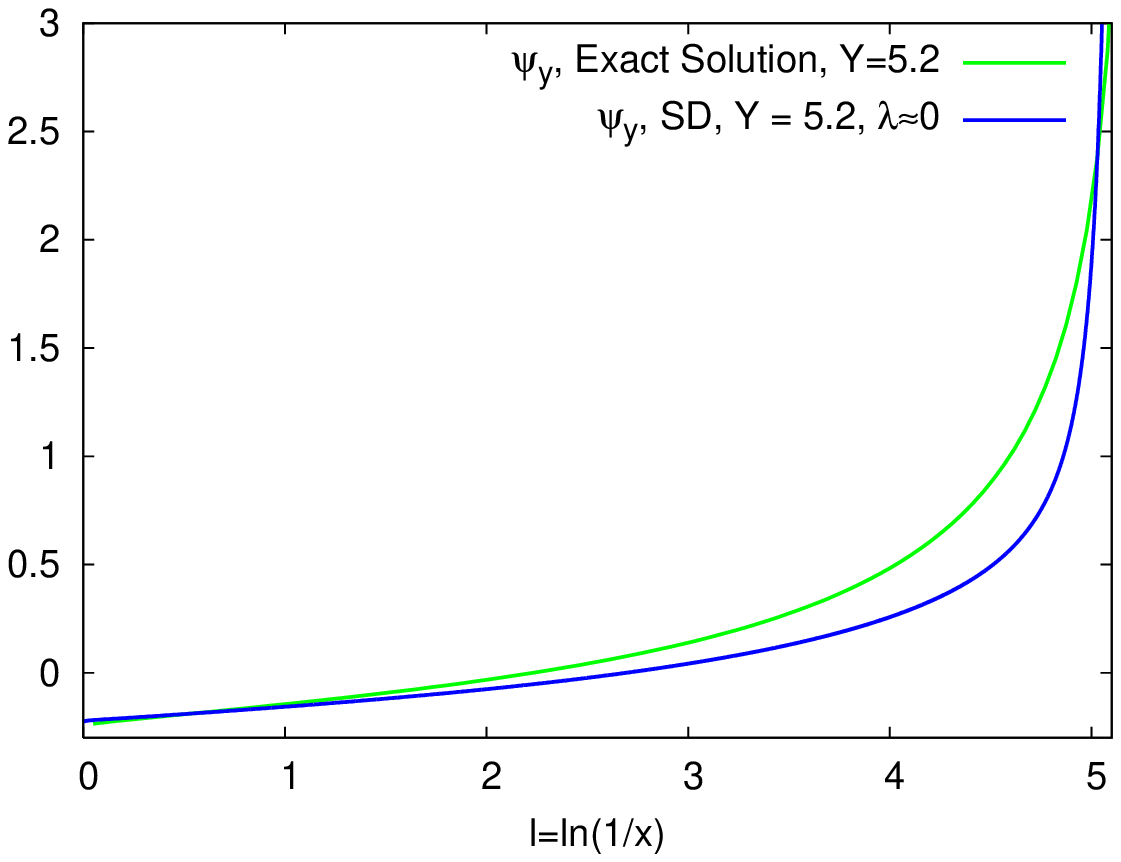, height=5truecm,width=0.48\tw}
\caption{\label{fig:MLLASTESpec} SD logarithmic derivatives $\psi_\ell$
(left) and $\psi_y$ (right) compared
with the ones of \cite{RPR2} at $Y=5.2$.}
\end{center}
\end{figure}

\begin{figure}[h]
\begin{center}
\epsfig{file=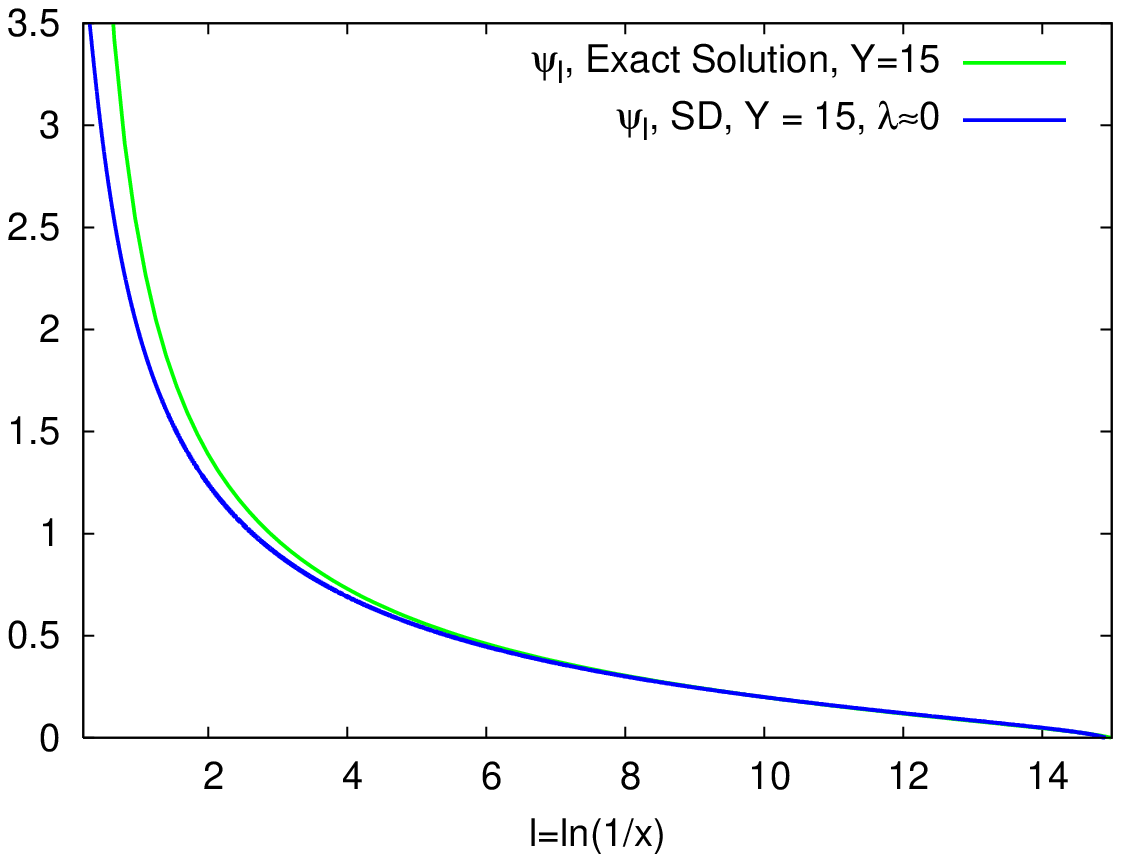, height=5truecm,width=0.48\tw}
\hfill
\epsfig{file=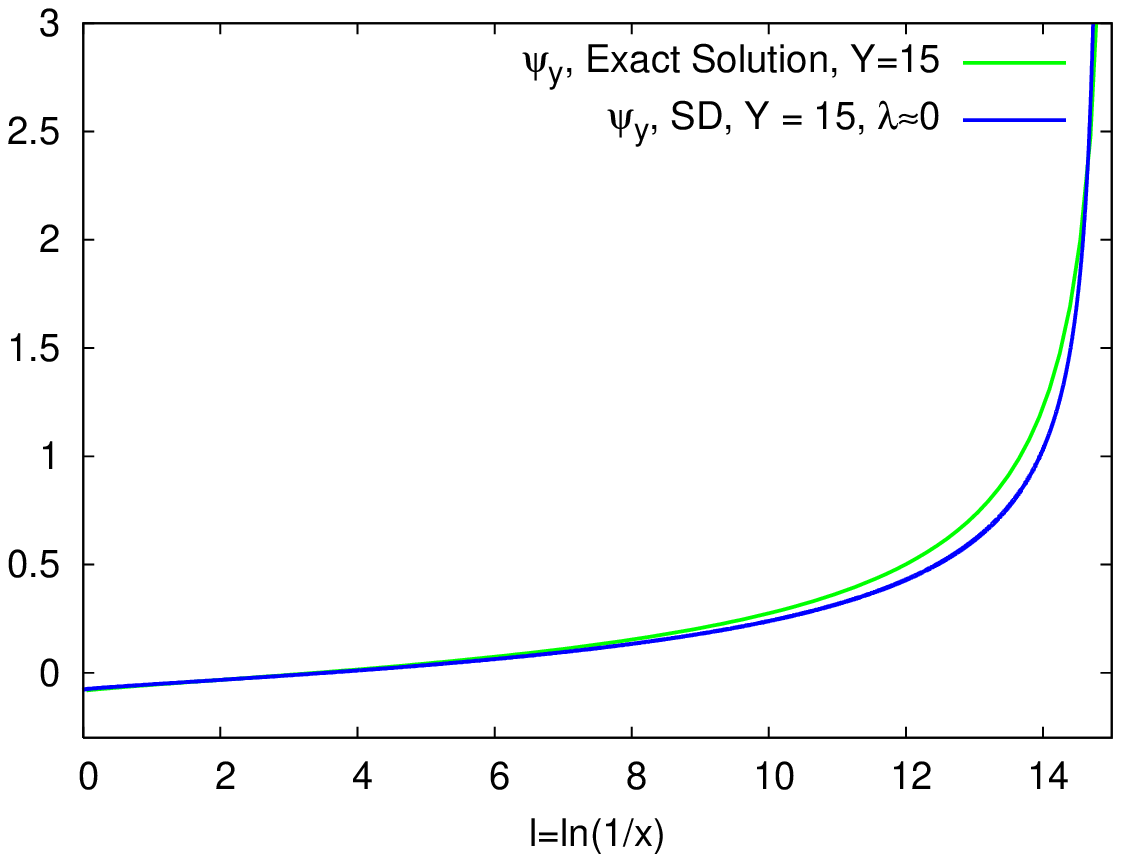, height=5truecm,width=0.48\tw}
\caption{\label{fig:MLLASTESpecbis} SD logarithmic derivatives $\psi_\ell$
(left) and $\psi_y$ (right) compared
with the ones of \cite{RPR2} at $Y=15$.}
\end{center}
\end{figure}

Since the logarithmic derivatives are ``infrared stable'' (see above),
we can take 
the limit $\lambda\to0$ in (\ref{eq:derpsi'l})(\ref{eq:derpsi'y})
\footnote{For this purpose, (\ref{eq:ratiomunu}) has been numerically
inverted.}, 
and compare their shapes with the
ones obtained in \cite{PerezMachet}; this is done
in  Figs.~\ref{fig:MLLASTESpec} and \ref{fig:MLLASTESpecbis}, at LEP-I energy
($E\Theta_0=91.2\,\text{GeV}$, $Y=5.2$) and at the unrealistic value $Y=15$.

The agreement between the SD and the exact logarithmic derivatives
is seen to be quite good.
The small deviations ($\leq 20\%$) that can be observed at large $\ell$
(the domain we deal with) arise from NMLLA
corrections that one does not control in the exact solution.
The agreement gets better and better as the energy increases.

It is checked in appendix (\ref{subsec:check}) that (\ref{eq:Specalphasrunmlla})
satisfies the evolution equation (\ref{eq:solg});
the SD logarithmic derivatives (\ref{eq:derpsi'l}) and (\ref{eq:derpsi'y}) can
therefore be used in the approximate calculation of
2-particle correlations at $\lambda\ne0$. This is what is done in the 
next section.

%%%%%%%%%%%%%%%%%%%%%%%%%%%%%%%%%%%%%%%%%%%%%%%%%%%%%
\section{2-PARTICLE CORRELATIONS INSIDE ONE JET AT 
$\boldsymbol{\lambda\!\not=\!0\ (Q_0\!\ne\!\Lambda_{QCD})}$}
\label{sec:TPC}
%%%%%%%%%%%%%%%%%%%%%%%%%%%%%%%%%%%%%%%%%%%%%%%%%%%%%

We study the correlation between 2-particles inside one jet of 
half opening angle $\Theta$ within the MLLA accuracy.
They have fixed energies $x_1=\omega_1/E$, $x_2=\omega_2/E$ ($\omega_1>\omega_2$)
and are emitted at arbitrary angles $\Theta_1$, $\Theta_2$. The constrain 
$\Theta_1\geq\Theta_2$ follows from the angular ordering in the cascading process. 
One has $\Theta\geq\Theta_1$ (see Fig.~1 of \cite{RPR2}).

\subsection{Variables and kinematics}
%%%%%%%%%%%%%%%%%%%%%%%%%%%%%%%%%%%%%

The variables and
kinematics of the cascading process are defined like in section 3.2 of \cite{RPR2}.

\subsection{MLLA evolution equations for correlations}
%%%%%%%%%%%%%%%%%%%%%%%%%%%%%%%%%%%%%%%%%%%%%%%%%%%%%%

The system of integral evolution equations 
for the quark and gluon jets two-particle 
correlation reads (see eqs.~(65) and (66) of \cite{RPR2})

\vbox{
\begin{eqnarray}
\label{eq:eveeqq}
\hskip -0.9cm Q^{(2)}(\ell_1,y_2,\eta)\!-\! Q_1(\ell_1,y_1)Q_2(\ell_2,y_2)
\!\!\!&\!\!=\!\!&\!\!\! \frac{C_F}{N_c}\!\!
\int_0^{\ell_1}\!\!\! d\ell\!\int_0^{y_2}\!\!\! dy\,
\gamma_0^2(\ell+y) \Big[\!1\!-\!\frac34 \delta(\ell-\ell_1) \!\Big]
G^{(2)}(\ell,y,\eta),\\\notag\\\notag\\
\notag
\hskip -0.9cm G^{(2)}(\ell_1,y_2,\eta) - G_1(\ell_1,y_1)G_2(\ell_2,y_2)
\!\!\!&\!\!\!=\!\!\!&\!\!\!\! 
\int_0^{\ell_1}\!\! d\ell\!\int_0^{y_2}\!\!dy\, \gamma_0^2(\ell+y)
\Big[\!1 - a \delta(\ell-\ell_1) \!\Big] G^{(2)}(\ell,y,\eta)\\\notag\\
\!\!\!&\!\!\!+\!\!\!&\!\!\! (a-b) \int_0^{y_2}dy \> \gamma_0^2(\ell_1+y)
G(\ell_1,y+\eta)G(\ell_1+\eta,y).
 \label{eq:eveeqglu} 
\end{eqnarray}
}

a is defined in (\ref{eq:adef}) while

\begin{equation}
b = \frac{1}{4N_c}\bigg[\frac{11}{3}N_c -\frac{4}{3}n_f T_R
\bigg(1-2\frac{C_F}{N_c}\bigg)^2\bigg]\stackrel{n_f=3}{=}0.915.
\label{eq:bdef}
\end{equation}

\subsection{MLLA solution at $\boldsymbol{\lambda \ne 0}$}
%%%%%%%%%%%%%%%%%%%%%%%%%%%%%%%%%%%%%%%%%%%%%%%%%%%%%%%%%%%%%%%%%%%%%

The quark and gluon jet correlators ${\cal C}_q$ and ${\cal C}_g$
have been exactly determined for any $\lambda$ in \cite{RPR2}
by respectively setting $Q^{(2)}={\cal C}_qQ_1Q_2$ and $G^{(2)}={\cal C}_gG_1G_2$
into (\ref{eq:eveeqq}) and (\ref{eq:eveeqglu}).
In the present work we limit ourselves to the exact MLLA solution which
consists in neglecting all ${\cal O}(\gamma_0^2)$ corrections in equations
(64) and (84) of \cite{RPR2}.

\subsubsection{Gluon jet}
%%%%%%%%%%%%%%%%%%%%%%%%%

At MLLA, the logarithmic derivatives of $\psi$ (\ref{eq:SP}) can be truncated to
the saddle point derivatives $\varphi_\ell,\,\varphi_y$ of (\ref{eq:phi}). The MLLA solution of (\ref{eq:eveeqglu}) then reads (see (77) in \cite{RPR2})

\begin{equation}
{\cal C}_g-1 \stackrel{MLLA}{\approx} \frac{1-b\left(\varphi_{1,\ell}
    +\varphi_{2,\ell} \right)-\delta_1} {1+ \bar\Delta + \Delta' + \delta_1}
\label{eq:CGMLLA}
\end{equation}

where we introduce
\begin{eqnarray}
&&\label{eq:deltabis}\bar\Delta = \gamma_0^{-2}
\Big(\varphi_{1,\ell}\varphi_{2,y}+\varphi_{1,y}\varphi_{2,\ell}\Big),
\\\notag\\
&&\label{eq:deltater}\Delta' = \gamma_0^{-2}
\Big(\varphi_{1,\ell}\delta\psi_{2,y}+\delta\psi_{1,y}\varphi_{2,\ell}+
\delta\psi_{1,\ell}\varphi_{2,y}+\varphi_{1,y}\delta\psi_{2,\ell}\Big);
\\\notag\\
&&\label{eq:chi} \chi=\ln\left(1+\frac1{1+\bar\Delta}\right),\quad 
\chi_{\ell}=\frac1{\chi}\frac{\partial\chi}{\partial\ell},\quad 
\chi_y=\frac1{\chi}\frac{\partial\chi}{\partial y};\\\notag\\
&&\delta_1 = \gamma_0^{-2}\Big[\chi_{\ell}(\varphi_{1,y}+\varphi_{2,y}) +
   \chi_{y}(\varphi_{1,\ell}+\varphi_{2,\ell})\Big].\label{eq:delta1}
\label{eq:nota4}
\end{eqnarray}

(\ref{eq:deltabis}) is obtained by using (\ref{eq:SPbis}):

\begin{equation}\label{eq:Delta}
\bar\Delta(\mu_1,\mu_2)=2\cosh(\mu_1-\mu_2)={\cal O}(1),
\end{equation}

which is the DLA contribution \cite{DLA}, while (\ref{eq:deltater}) 
(see appendix \ref{subsec:Deltaprime})
is obtained by using (\ref{eq:SPbis}), (\ref{eq:derpsiprime1}) and 
(\ref{eq:derpsiprime1bis})

\begin{equation}\label{eq:Deltaprime}
\Delta'(\mu_1,\mu_2)=\frac{ e^{-\mu_1}\delta\psi_{2,\ell} + e^{-\mu_2}\delta\psi_{1,\ell} 
        + e^{\mu_1}\delta\psi_{2,y} + e^{\mu_2}\delta\psi_{1,y}}
{\gamma_0}={\cal O}(\gamma_0);
\end{equation}

it is a next-to-leading (MLLA) correction.
To get (\ref{eq:chi}), we first use (\ref{eq:Delta}), which gives
\begin{eqnarray}\label{eq:chielly}
\chi_\ell =  -\frac{\tanh{\frac{\mu_1-\mu_2}{2}}}{1+2\cosh(\mu_1\!-\!\mu_2)}
 \left(\frac{\partial\mu_1}{\partial\ell} -\frac{\partial\mu_2}{\partial\ell}
 \right),\quad
 \chi_y = -\frac{\tanh{\frac{\mu_1-\mu_2}{2}}}{1+2\cosh(\mu_1\!-\!\mu_2)}
 \left(\frac{\partial\mu_1}{\partial y} -\frac{\partial\mu_2}{\partial y} \right),
\end{eqnarray}

and then (\ref{eq:dermul1}) to get
\begin{eqnarray*}
  \chi_\ell\!=\! \beta\gamma_0^2\,
  \frac{\tanh{\frac{\mu_1-\mu_2}{2}}}{1\!+\!2\cosh(\mu_1\!-\!\mu_2)} \>
  \frac{e^{\mu_1}\widetilde{Q}_1-e^{\mu_2}\widetilde{Q}_2}{2},\quad
 \chi_y \!=\! -\beta\gamma_0^2
  \frac{\tanh{\frac{\mu_1-\mu_2}{2}}}{1\!+\!2\cosh(\mu_1\!-\!\mu_2)} \>
  \frac{e^{-\mu_1}\widetilde{Q}_1-e^{-\mu_2}\widetilde{Q}_2}{2}
\end{eqnarray*}

which are ${\cal O}(\gamma_0^2)$. They  should be plugged into
(\ref{eq:nota4}) together with (\ref{eq:SPbis}), which gives

\begin{equation}\label{eq:delta2}
 \delta_1 = \beta\gamma_0
\frac{2\sinh^2{\left(\frac{\mu_1-\mu_2}2\right)}}
{3+4\sinh^2{\left(\frac{\mu_1-\mu_2}2\right)}}
\Big(\widetilde Q(\mu_1,\upsilon_1)+
\widetilde Q(\mu_2,\upsilon_2)\Big)={\cal O}(\gamma_0);
\end{equation}

it is also a MLLA term. For $Q\gg Q_0\geq\Lambda_{QCD}$ we finally get,
\begin{equation}
{\cal C}_g(\ell_1,\ell_2,Y,\lambda)\stackrel{MLLA}{\approx}
1 + \frac{1 - b\gamma_0\left(e^{\mu_1} + e^{\mu_2} \right) - \delta_1} 
{1+ 2\cosh(\mu_1-\mu_2) + \Delta'(\mu_1,\mu_2) + \delta_1}
\label{eq:CGMLLAbis}
\end{equation}

where the expression for $\Delta'$ (\ref{eq:dDelta}) 
is written in appendix \ref{subsec:Deltaprime}.
It is important to notice that $\delta_1\simeq0$ near $\ell_1\approx\ell_2$ ($\mu_1\approx\mu_2$) while it 
is positive and increases as $\eta$ 
gets larger (see (\ref{eq:tildeQ}) and Fig.\ref{fig:tildeQ});
it makes the correlation function narrower in $|\ell_1-\ell_2|$.

\subsubsection{Quark jet}
%%%%%%%%%%%%%%%%%%%%%%%%%

The MLLA solution of (\ref{eq:eveeqq}) reads (see (93) in \cite{RPR2})
\begin{eqnarray}
\frac{{\cal C}_q-1}{{\cal C}_g-1} \stackrel{MLLA}{\approx}
\frac{N_c}{C_F}\bigg[1+(b-a)(\phi_{1,\ell} + \phi_{2,\ell})
\frac{1+\bar\Delta}{2+\bar\Delta}\bigg]
\label{eq:rapMLLA}
\end{eqnarray}

Inserting (\ref{eq:deltabis})-(\ref{eq:nota4}) into (\ref{eq:rapMLLA})  we get
\begin{equation*}
{\cal C}_q(\ell_1, \ell_2, Y, \lambda)\stackrel{MLLA}{\approx}1+
\frac{N_c}{C_F}\Big({\cal C}_g(\ell_1, \ell_2, Y, \lambda)-1\Big)
\bigg[1+(b-a)\gamma_0(e^{\mu_1} + e^{\mu_2})
\frac{1+2\cosh(\mu_1-\mu_2)}{2+2\cosh(\mu_1-\mu_2)}\bigg].
\end{equation*}

which finally reduces (for $Q\gg Q_0\geq\Lambda_{QCD}$) to
\begin{equation}\label{eq:rapMLLAbis}
{\cal C}_q(\ell_1,\ell_2, Y, \lambda)\stackrel{MLLA}{\approx}1+
\frac{N_c}{C_F}
\bigg[\Big({\cal C}_g(\ell_1, \ell_2, Y, \lambda)-1\Big)+\frac12(b-a)\gamma_0\frac{e^{\mu_1} + e^{\mu_2}}
{1+\cosh(\mu_1-\mu_2)}\bigg].
\end{equation}

\subsection{Sensitivity of the quark and gluon jets correlators to the value
of $\boldsymbol{\lambda}$}
%%%%%%%%%%%%%%%%%%%%%%%%%%%%%%%%%%%%%%%%%%%%%%%%%%%%%%%%%%%%%%%%%%%%%%%%%%%%

Increasing $\lambda$ translates into taking the limits $\beta,\,\Lambda_{QCD}\to0$
($Y=\ell+y\ll\lambda,\, Q\gg Q_0\gg\Lambda_{QCD}$) in the definition of the anomalous 
dimension via the running coupling constant ($\gamma_0=\gamma_0(\alpha_s)$, see (44) in 
\cite{RPR2}). It allows to neglect $\ell$, $y$ with respect to $\lambda$ as follows
\begin{equation}\label{eq:gammaconst}
\gamma_0^2(\ell+y)=\frac1{\beta(\ell+y+\lambda)}
\stackrel{\ell+y\ll\lambda}{\approx}\gamma_0^2=\frac1{\beta\lambda},
\end{equation}

such that $\gamma_0$ can be taken as a constant. Estimating (\ref{eq:MLLAalphasrun}) 
in the region $\lambda\gg1\Leftrightarrow s\ll1$ needs evaluating the kernel
\begin{eqnarray}
\frac1{\nu+s}\left(\frac{\omega\left(\nu+s\right)}
{\left(\omega+s\right)\nu}\right)^{1/\beta\left(\omega-\nu\right)}\!\!
\left(\frac{\nu}{\nu+s}\right)^{a/\beta}
\!\!&\!\!\stackrel{s\ll1}{\approx}\!\!&\!\!
\frac1{\nu}\left(1+\frac{\omega-\nu}
{\omega\nu}s\right)^{1/\beta\left(\omega-\nu\right)}\left(1-\frac{s}{\nu}\right)^{a/\beta}
\cr\cr
&&\hskip -7cm
\approx\frac1{\nu}\left[1+\frac1{\nu}\left(\frac1\omega-a\right)\frac{s}\beta+\frac1{2!}
\frac1{\nu^2}\left(\frac1\omega-a\right)^2\frac{s^2}{\beta^2}+\frac1{3!}\frac1{\nu^3}
\left(\frac1\omega-a\right)^3\frac{s^3}{\beta^3}+\dots\right].\label{eq:MLLApropag}
\end{eqnarray}

Integrating (\ref{eq:MLLApropag}) over $s$, using (\ref{eq:gammaconst}) and
$\int_0^{\infty}s^n\,e^{-\lambda s}=\frac{n!}{\lambda^n}$, we get
\begin{eqnarray*}
{\cal {G}}(\omega,\nu)
\!\!&\!\!\approx\!\!&\!\!\frac1{\nu}
\left[1+\frac1{\nu}\left(\frac1\omega-a\right)\frac1{\beta\lambda}+
\frac1{\nu^2}\left(\frac1\omega-a\right)^2\left(\frac1{\beta\lambda}\right)^2+
\frac1{\nu^3}\left(\frac1\omega-a\right)^3
\left(\frac1{\beta\lambda}\right)^3+\dots\right]\cr\cr
\!\!&\!\!=\!\!&\!\!\frac1{\nu-\gamma_0^2\left(1/\omega-a\right)},
\end{eqnarray*}

which, after inverting the Mellin's representation (132) of \cite{RPR2}, gives

\begin{equation}\label{eq:Gexpres}
G(\ell,y)\stackrel{x\ll1}{\simeq}\exp(2\gamma_0\sqrt{\ell\,y}-a\gamma_0^2y).
\end{equation}

Taking the same limit in (\ref{eq:ratiomunu}) and (\ref{eq:relmunu}) gives
respectively

\begin{equation}\label{eq:tanhmu}
\frac{y-\ell}{y+\ell}\stackrel{\ell+y\ll\lambda}\approx\tanh\mu\Rightarrow
\mu=\frac12\ln\frac{y}{\ell},\qquad
\mu-\upsilon\stackrel{\ell+y\ll\lambda}{\approx}\frac12\frac{y-\ell}{\lambda}
\Rightarrow\mu\sim\upsilon.
\end{equation}

Furthermore, we use (\ref{eq:tanhmu}) to show how (\ref{eq:SP}) reduces to
the exponent in (\ref{eq:Gexpres})
\footnote{we set $\beta=0$ in (\ref{eq:derpsi'l}), (\ref{eq:derpsi'y}) 
and only consider terms $\propto a$}
\begin{eqnarray}
\phi\!\!&\!\!=\!\!&\!\!\frac2{\sqrt{\beta}}\frac{\mu-\upsilon}{\sinh\mu-\sinh\upsilon}
\stackrel{\ell+y\ll\lambda}{\approx}
2\gamma_0\sqrt{\ell\, y},\cr\cr\cr
\left(\frac{\nu_0}{\nu_0+s_0}\right)^{a/\beta}\!\!&\!\!=\!\!&\!\!
-\frac12\frac{a}\beta\ln\left(1+\frac{\ell+y}\lambda\right)-
\frac{a}{\beta}(\mu-\upsilon)\approx-\frac12\frac{a}\beta\frac{\ell+y}\lambda-
\frac{a}{\beta}(\mu-\upsilon)\cr\cr
\!\!&\!\!\stackrel{\ell+y\ll\lambda}{\approx}\!\!&\!\!-a\gamma_0^2\,y.
\end{eqnarray}

Thus, since $\mu=\frac12\ln\frac{y}{\ell}$ (\ref{eq:tanhmu}),
(\ref{eq:derpsi'l}) and (\ref{eq:derpsi'y}) simplify to

\begin{equation}
\psi_\ell\stackrel{\ell+y\ll\lambda}{\approx}\gamma_0e^{\mu}=\gamma_0\sqrt{\frac{y}{\ell}},
\qquad\psi_y\stackrel{\ell+y\ll\lambda}{\approx}\gamma_0e^{-\mu}-a\gamma_0^2=
\gamma_0\sqrt{\frac{\ell}{y}}-a\gamma_0^2.
\end{equation}

Therefore, taking the limit $\beta,\,\Lambda_{QCD}\to0$ ($\lambda\to\infty$)
leads to the simplified model
described in section 4.2 of \cite{RPR2}. Setting, for the sake of simplicity,
$\ell_1\approx\ell_2$ in (\ref{eq:CGMLLAbis})(\ref{eq:rapMLLA}),
where $\delta_1$ vanishes, we obtain, in the high energy limit
\begin{equation}\label{eq:cgqmodel}
{\cal C}_g(\ell,y)\simeq1\!+\!\frac13\left[1\!-\!2\left(b\!-\!\frac13a\right)
\psi_{\ell}(\ell,y)\right],\quad
{\cal C}_q(\ell,y)\simeq1\!+\!\frac{N_c}{C_F}\left[\frac13\!-\!\frac12
\left(\frac53a\!+\!b\right)\psi_{\ell}(\ell,y)\right],
\end{equation}
where
\begin{eqnarray}
&&b-\frac13 a = \frac1{18}\left(11-8\frac{T_R}{N_c}
 +28\frac{T_R}{N_c}\frac{C_F}{N_c}-24\frac{T_R}{N_c}
\frac{C_F^2}{N_c^2}\right)\stackrel{n_f=3}{\approx}0.6,\cr\cr\cr
&&\frac53 a + b = \frac29\left(11 + \frac{T_R}{N_c}
 +\frac{T_R}{N_c}\frac{C_F}{N_c} -6\frac{T_R}{N_c}
\frac{C_F^2}{N_c^2}\right)\stackrel{n_f=3}{\approx}2.5.
\end{eqnarray}

\begin{figure}[h]
\begin{center}
\epsfig{file=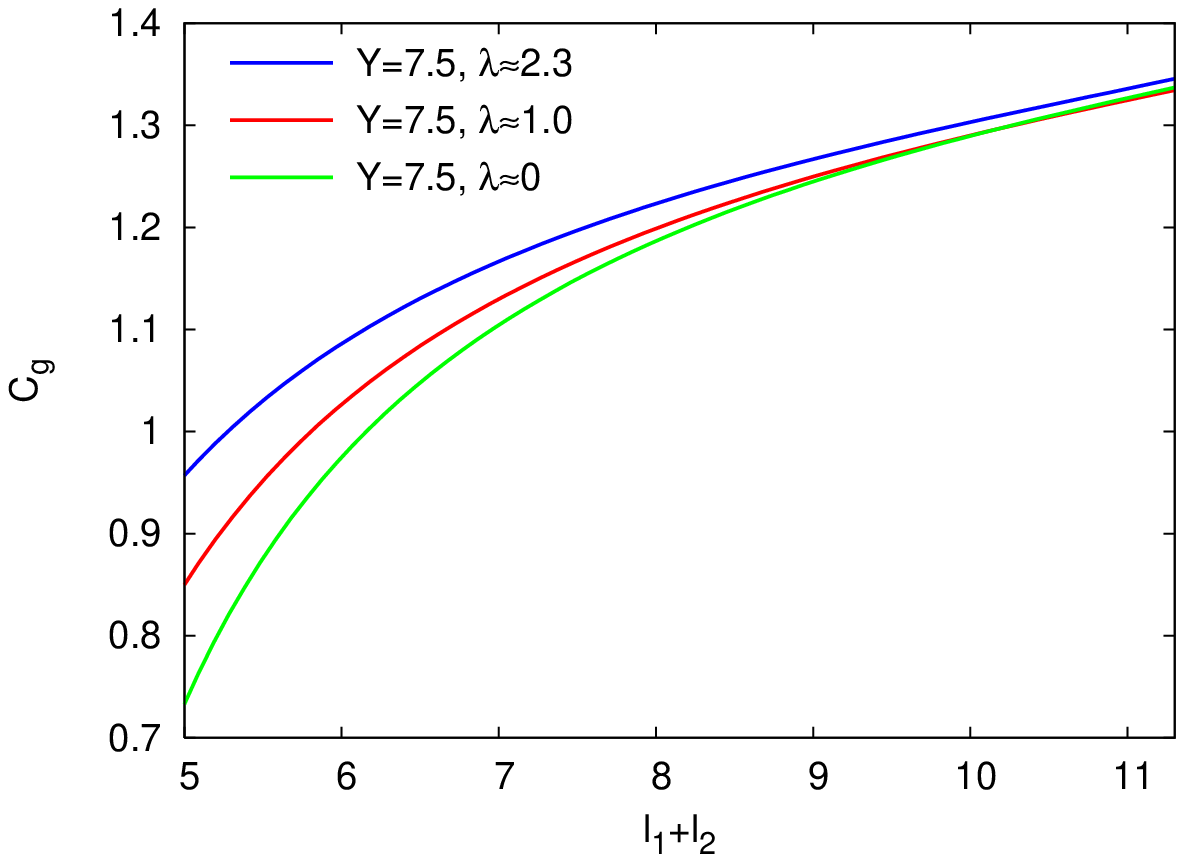, height=5truecm,width=0.48\tw}
\hfill
\epsfig{file=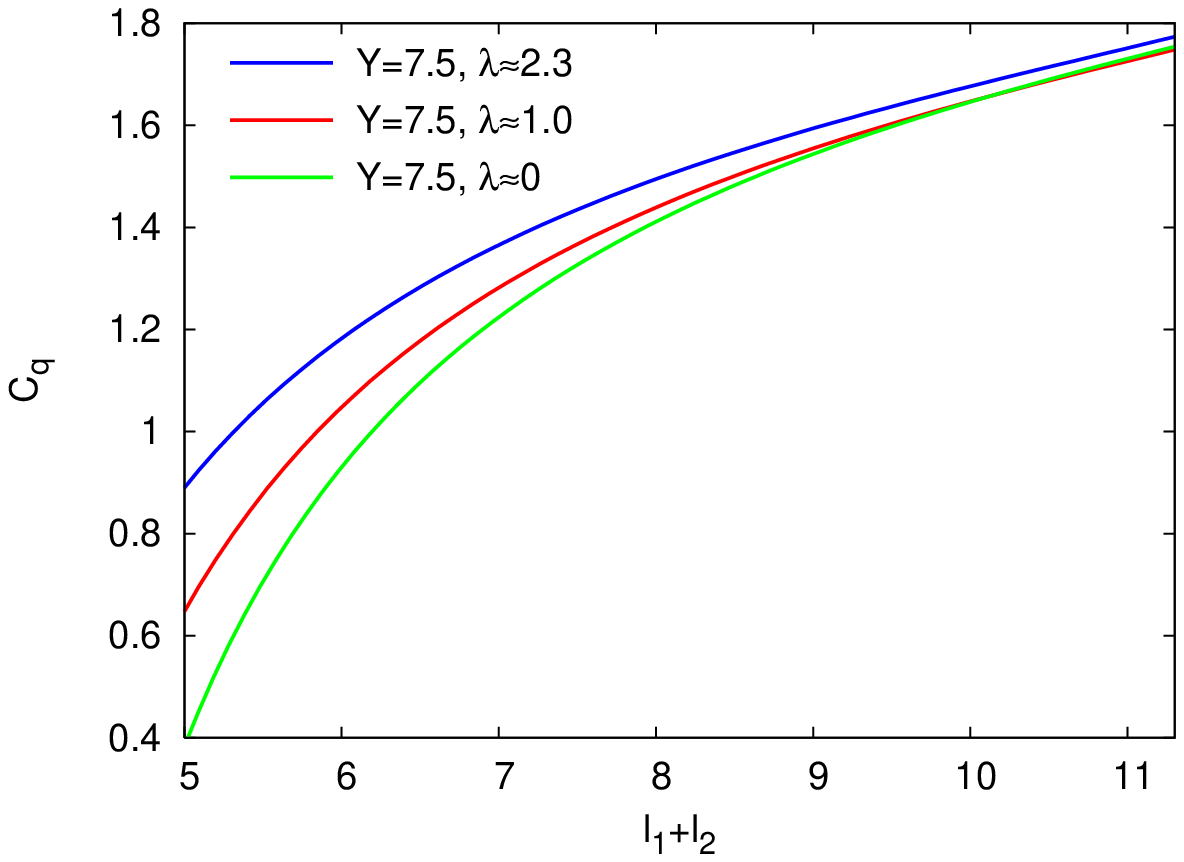, height=5truecm,width=0.48\tw}
\caption{\label{fig:LHCcorrg} Varying $\lambda$ at fixed $Q_0$;
$\Lambda_{QCD}$ dependence of  ${\cal C}_g$ (left) and ${\cal C}_q$ (right)}
\end{center}
\end{figure}

Thus, when $\lambda$ increases by decreasing $\Lambda_{QCD}$, 
$\psi_\ell\!\propto\!\gamma_0$ decreases and
the correlators (\ref{eq:cgqmodel}) increase.
For LHC, a typical value is $Y=7.5$ and we compare
in Fig.~\ref{fig:LHCcorrg}, at fixed $Q_0$,
 the limiting case $\lambda\approx0$
($Q_0\approx\Lambda_{QCD}\approx 253\,\text{MeV}$) with
$\lambda\approx1.0$ ($\Lambda_{QCD}=100\,\text{MeV}$) and $\lambda \approx
2.3$ ($\Lambda_{QCD}=25\,\text{MeV}$). As predicted by (\ref{eq:cgqmodel}),
the correlation increases when $\Lambda_{QCD}\to 0$ at fixed $Q_0$.

\begin{figure}
\vbox{
\begin{center}
\epsfig{file=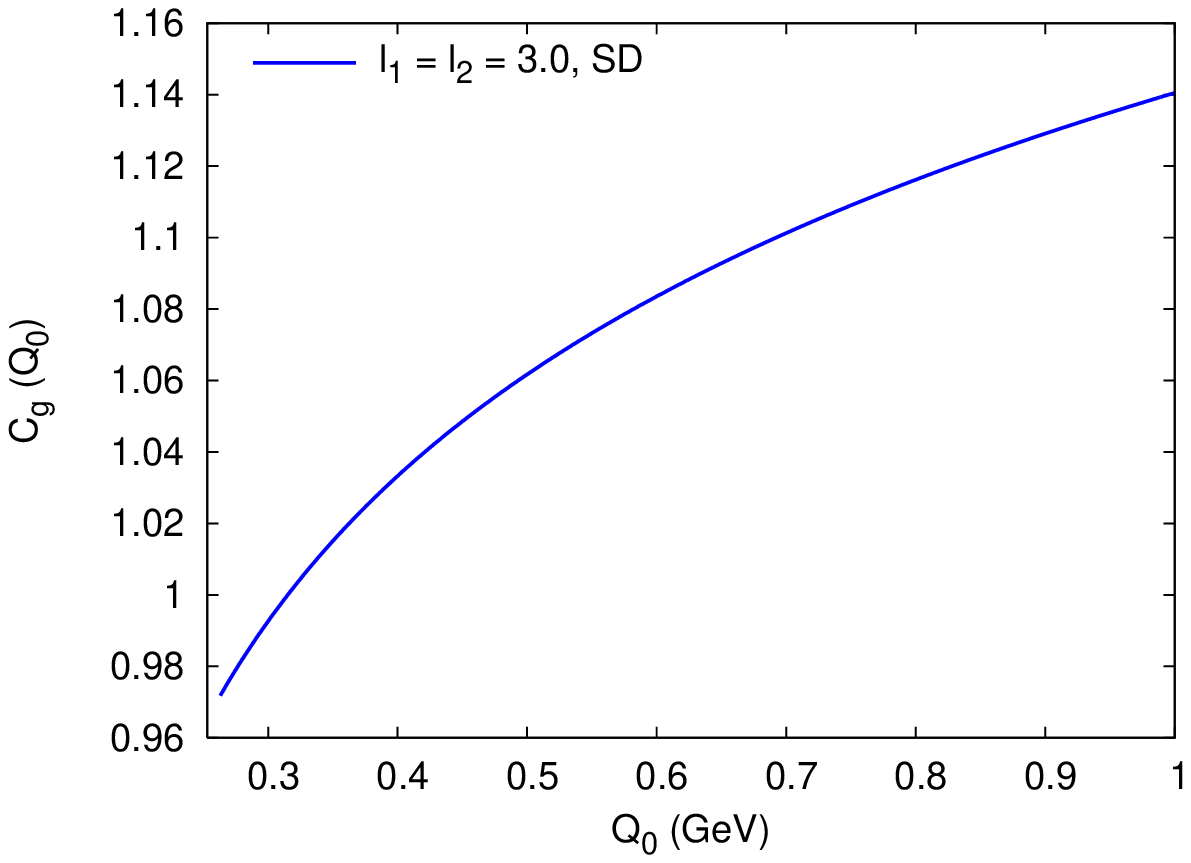, height=5truecm,width=0.48\tw}
\hfill
\epsfig{file=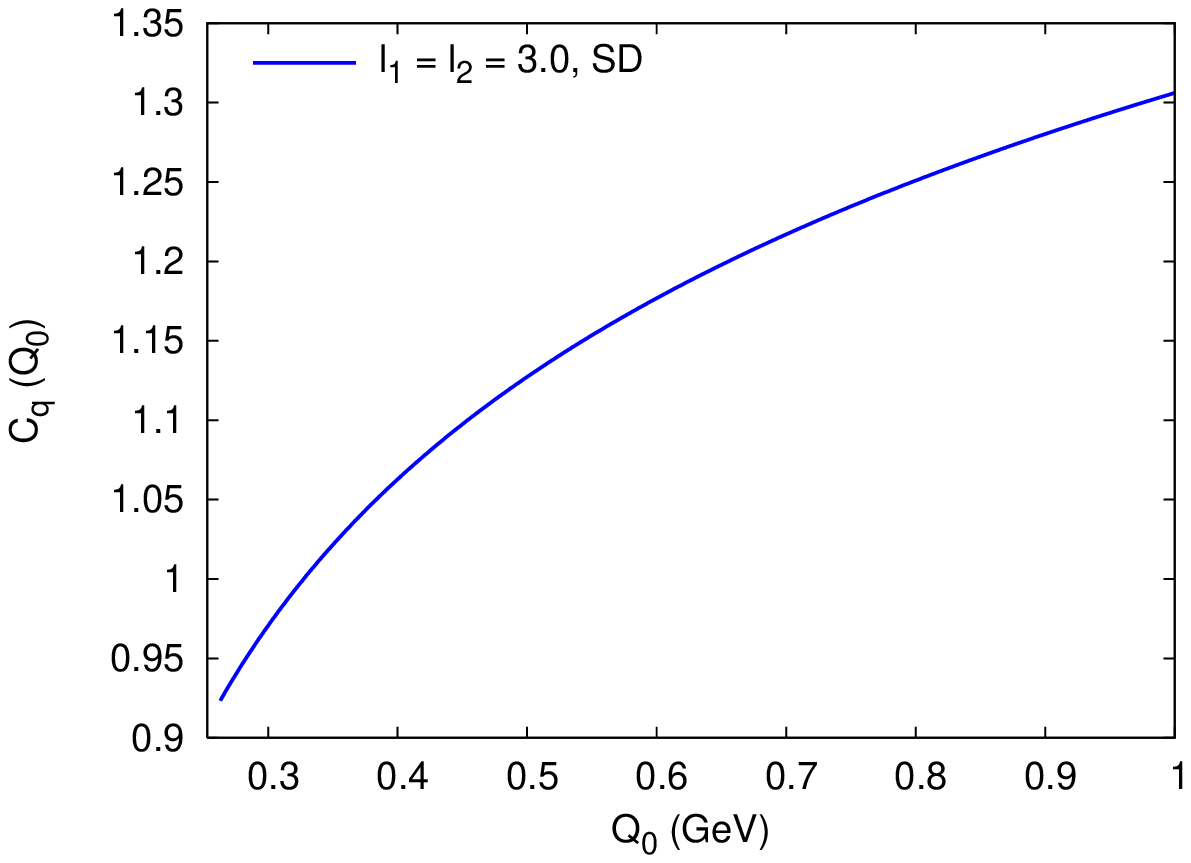, height=5truecm,width=0.48\tw}
\caption{Varying $\lambda$ at fixed $\Lambda_{QCD}=253\,MeV$; $Q_0$-dependence
of ${\cal C}_g$ (left) and ${\cal C}_q$ (right) 
at $\ell_1=\ell_2=3.0$}
\label{fig:Qeff}
\end{center}
}
\end{figure}

It is also sensitive to the value of $Q_0$. As
seen in (\ref{eq:cgqmodel}), since $y=\ln\frac{Q}{Q_0}-\ell$, if one increases
$Q_0$ (since $\Lambda_{QCD}$ is fixed, $\gamma_0$ does not change), 
thereby reducing the available phase space, the correlators increase.
This dependence of the correlators at fixed $\Lambda_{QCD}$
is displayed in Fig.\ref{fig:Qeff} for $0.3\,\text{GeV}\leq Q_0\leq1.0\,\text{GeV}$
at $\ell_1=\ell_2=3.0$ (soft parton).

In the simplified model which leads to (\ref{eq:cgqmodel}),
 ${\cal C}_g$ and ${\cal C}_q$
respectively go to the asymptotic values $4/3$ and $1 + N_c/3C_F$. This is
however not the case in the general situation  $\beta \not=0$, as can
be easily checked by using (\ref{eq:derpsi'l}) and (\ref{eq:derpsi'y}); 
for example, near the maximum
of the distribution ($\mu \sim v \to 0$), a contribution 
$\propto\lambda^{3/2}/[(Y+\lambda)^{3/2}-\lambda^{3/2}]$ occurs in
the term proportional to $\beta$ in (\ref{eq:cgqmodel}) that
yields negative values of $\psi_\ell$ when $\lambda$ increases.

\subsection{Extension of the Fong and Webber expansion;
its limit $\boldsymbol{\lambda=0}$}
%%%%%%%%%%%%%%%%%%%%%%%%%%%%%%%%%%%%%%%%%%%%%%%%%%%%%%%%

In the Fong-Webber regime, the energies of the two registered particles
stay very close to the peak of the inclusive hump-backed distribution that
is, $|\ell_i-\ell_{max}|\ll \sigma \propto [(Y+\lambda)^{3/2}-\lambda^{3/2}]^{1/2}$ 
(see (\ref{eq:gaussian})).

Near the maximum of the single inclusive distribution $\ell_1\sim\ell_2\simeq Y/2$ 
($\mu,\upsilon\to0$, see appendix \ref{subsec:DeTA})

$$
\lim_{\mu,\upsilon\rightarrow0}C=\left(\frac{\lambda}{Y+\lambda}\right)^{1/2}
,\quad\lim_{\mu,\upsilon\rightarrow0}K_i=\frac32\frac{\upsilon^2_i}{\mu^3_i-\upsilon^3_i},
\quad\lim_{\mu,\upsilon\rightarrow0}\widetilde{Q}=
\frac23+\frac13\left(\frac{\lambda}{Y+\lambda}\right)^{3/2},
$$

where $C$, $K_i$ and $\tilde Q$ are defined in (\ref{eq:CC}), (\ref{eq:KK}) and (\ref{eq:tildeQ}).
Keeping only the  terms linear in $\mu$ and the term quadratic in the difference 
$(\mu_1-\mu_2)$, one has
\begin{eqnarray}
\bar\Delta+\Delta'\stackrel{\ell_1\sim\ell_2\simeq Y/2}{\simeq}
2+(\mu_1-\mu_2)^2-a\gamma_0\left(2+\mu_1+\mu_2\right)-\beta\gamma_0\left[2
+3\frac{\lambda^{3/2}}{(Y+\lambda)^{3/2}-\lambda^{3/2}}\right]
\end{eqnarray}

and
\begin{equation}
\delta_1\stackrel{\ell_1\sim\ell_2\simeq Y/2}{\simeq}\frac19\beta\gamma_0
(\mu_1-\mu_2)^2\left[2+\left(\frac{\lambda}{Y+\lambda}\right)^{3/2}\right];
\end{equation}

$\delta_1$ can be neglected, since $\gamma_0(\mu_1-\mu_2)^2\ll(\mu_1-\mu_2)^2\ll1$.
Then, in the same limit, (\ref{eq:CGMLLAbis}), (\ref{eq:rapMLLAbis}) become
\begin{equation}\label{eq:FWgluon}
{\cal C}_g^0(\ell_1,\ell_2,Y,\lambda)\!\stackrel{\ell_1\sim\ell_2\simeq Y/2}{\simeq}\!1\!+\!\frac{1-b\gamma_0(2+\mu_1+\mu_2)}
{3+(\mu_1-\mu_2)^2-a\gamma_0\left(2+\mu_1+\mu_2\right)-\beta\gamma_0\left[2
+3\displaystyle\frac{\lambda^{3/2}}{(Y+\lambda)^{3/2}-\lambda^{3/2}}\right]},
\end{equation}

\begin{equation}\label{eq:FWquark}
{\cal C}_q^0(\ell_1, \ell_2, Y, \lambda)\stackrel{\ell_1\sim\ell_2\simeq Y/2}{\simeq}
1+\frac{N_c}{C_F}
\bigg[\Big({\cal C}_g^0(\ell_1, \ell_2, Y, \lambda)-1\Big)+
\frac14(b-a)\gamma_0\left(2+\mu_1+\mu_2\right)\bigg].
\end{equation}

Using (\ref{eq:ellmu}) one has
$$
(\mu_1\!-\!\mu_2)^2\simeq9\frac{Y+\lambda}
{\left[(Y+\lambda)^{3/2}-\lambda^{3/2}\right]^2}(\ell_1-\ell_2)^2,
\quad\mu_1\!+\!\mu_2\simeq3\frac{(Y+\lambda)^{1/2}}{(Y+\lambda)^{3/2}-\lambda^{3/2}}
\left[Y\!-\!(\ell_1+\ell_2)\right]
$$
such that the expansion of (\ref{eq:FWgluon}), (\ref{eq:FWquark}) 
in $\gamma_0\propto\sqrt{\alpha_s}$ reads

\vbox{
\begin{eqnarray}
{\cal C}_g^0(\ell_1,\ell_2,Y,\lambda)\!\!&\!\!\simeq\!\!&\!\!\frac43-
\left(\!\frac{(Y+\lambda)^{1/2}(\ell_1-\ell_2)}
{(Y+\lambda)^{3/2}-\lambda^{3/2}}\!\right)^2+
\left(\!\frac23+\frac{\left(Y+\lambda\!\right)^{1/2}Y}
{(Y+\lambda)^{3/2}-\lambda^{3/2}}\right)\!\!\left(\!\frac13a-b\!\right)\gamma_0\cr\cr
&&\hskip -2.5cm+\frac13\left(\frac23+\frac{\lambda^{3/2}}{(Y+\lambda)^{3/2}-\lambda^{3/2}}\right)
\beta\gamma_0+\left(\!b-\frac13a\!\right)\left(\frac{\left(Y+\lambda\right)^{1/2}(\ell_1+\ell_2)}
{(Y+\lambda)^{3/2}-\lambda^{3/2}}\right)\gamma_0+{\cal O}(\gamma_0^2),
\end{eqnarray}
}
\vbox{
\begin{eqnarray}
{\cal C}_q^0(\ell_1,\ell_2,Y,\lambda)\!\!&\!\!\simeq\!\!&\!\!1\!+\!\frac{N_c}{3C_F}\!+\!
\frac{N_c}{C_F}\Bigg[\!\!-\!
\left(\frac{(Y+\lambda)^{1/2}(\ell_1-\ell_2)}
{(Y+\lambda)^{3/2}-\lambda^{3/2}}\right)^2\!\!\!\!-\frac14\!\!
\left(\!\frac23\!+\!\frac{\left(Y+\lambda\right)^{1/2}Y}
{(Y+\lambda)^{3/2}-\lambda^{3/2}}\!\right)\!\!
\left(\!\frac53a\!+\!b\!\right)\gamma_0\bigg.\cr\cr&&\hskip -2.5cm\bigg.+\frac13\left(\frac23+\frac{\lambda^{3/2}}{(Y+\lambda)^{3/2}-\lambda^{3/2}}\right)
\beta\gamma_0+\frac14\left(\!\frac53a+b\!\right)
\left(\frac{\left(Y+\lambda\right)^{1/2}(\ell_1+\ell_2)}
{(Y+\lambda)^{3/2}-\lambda^{3/2}}\right)\gamma_0\Bigg]+{\cal O}(\gamma_0^2).
\end{eqnarray}
}
Therefore, near the hump of the single inclusive distribution,
(\ref{eq:CGMLLAbis}),(\ref{eq:rapMLLAbis}) behave as a linear functions
of the sum $(\ell_1+\ell_2)$ and as a quadratic functions of the difference
$(\ell_1-\ell_2)$. At the limit $\lambda=0$, one recovers the Fong-Webber expression 
\cite{FW}.

\subsection{Comparison with the exact solution of the evolution equations:
$\boldsymbol{\lambda=0}$}
%%%%%%%%%%%%%%%%%%%%%%%%%%%%%%%%%%%%%%%%%%%%%%%%%%%%%%%%%%%%%%%%%%%%%%%%%%

\begin{figure}[h]
\begin{center}
\epsfig{file=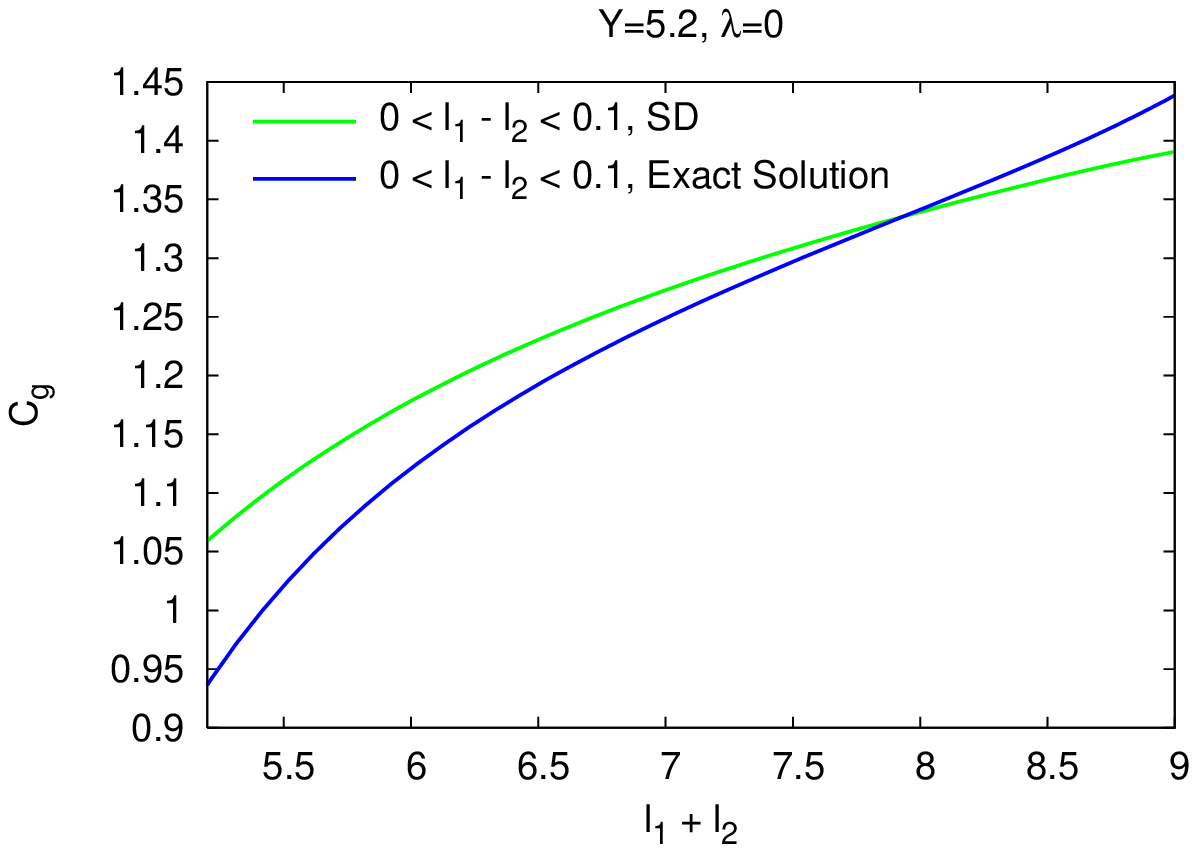, height=5truecm,width=0.48\tw}
\hfill
\epsfig{file=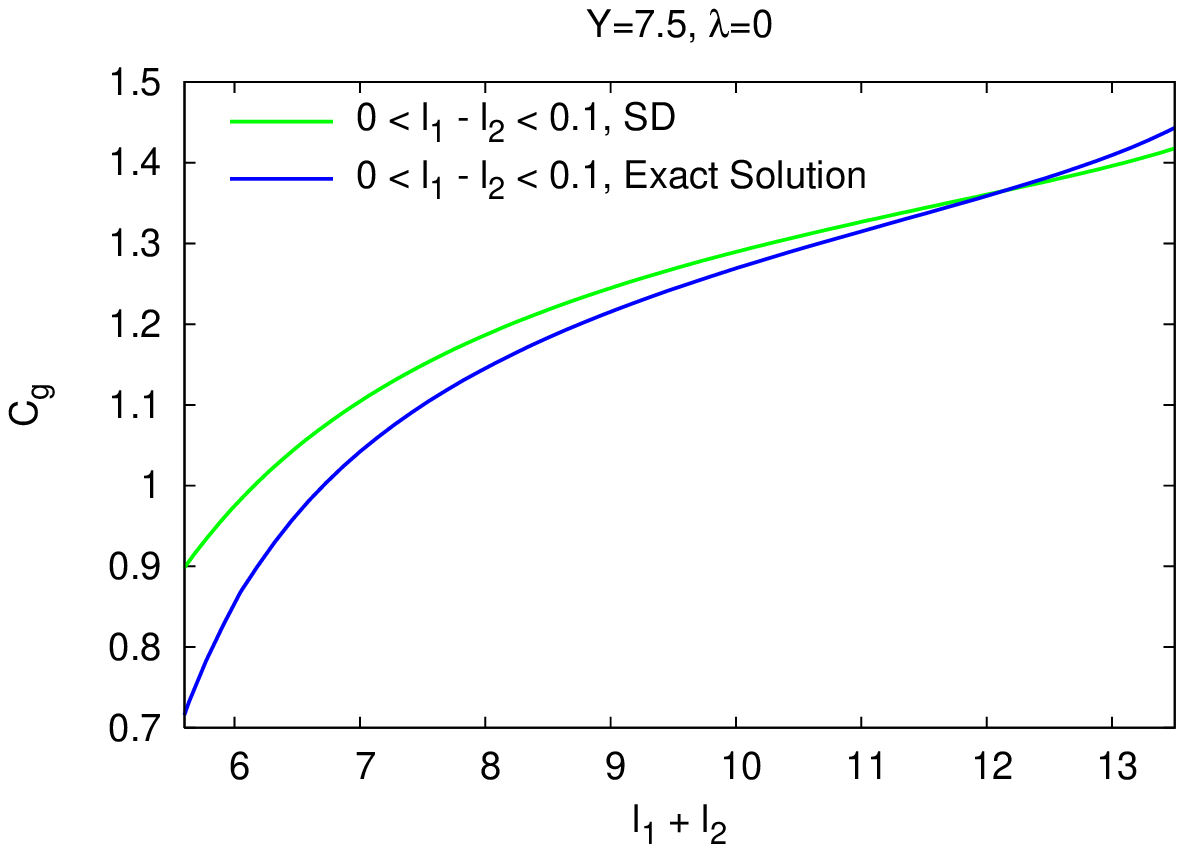, height=5truecm,width=0.48\tw}
\vskip 0.5cm
\epsfig{file=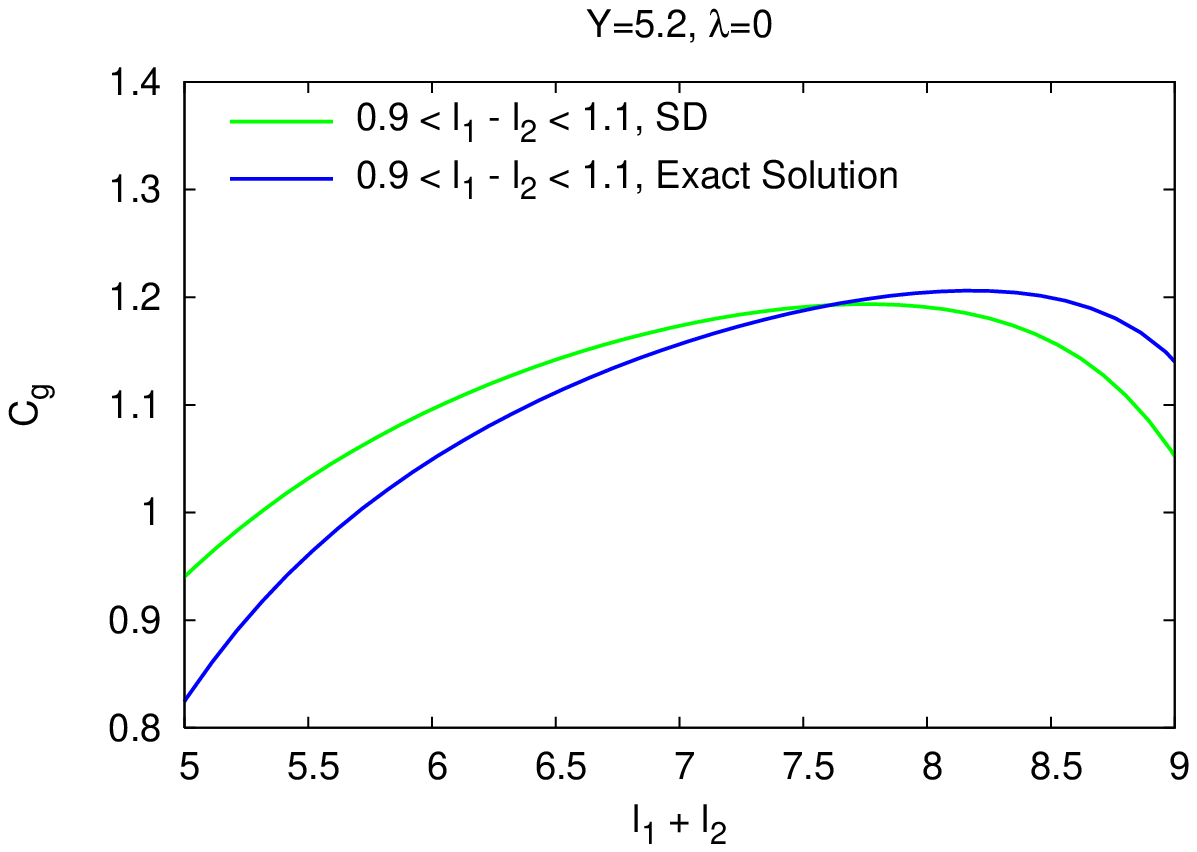, height=5truecm,width=0.48\tw}
\hfill
\epsfig{file=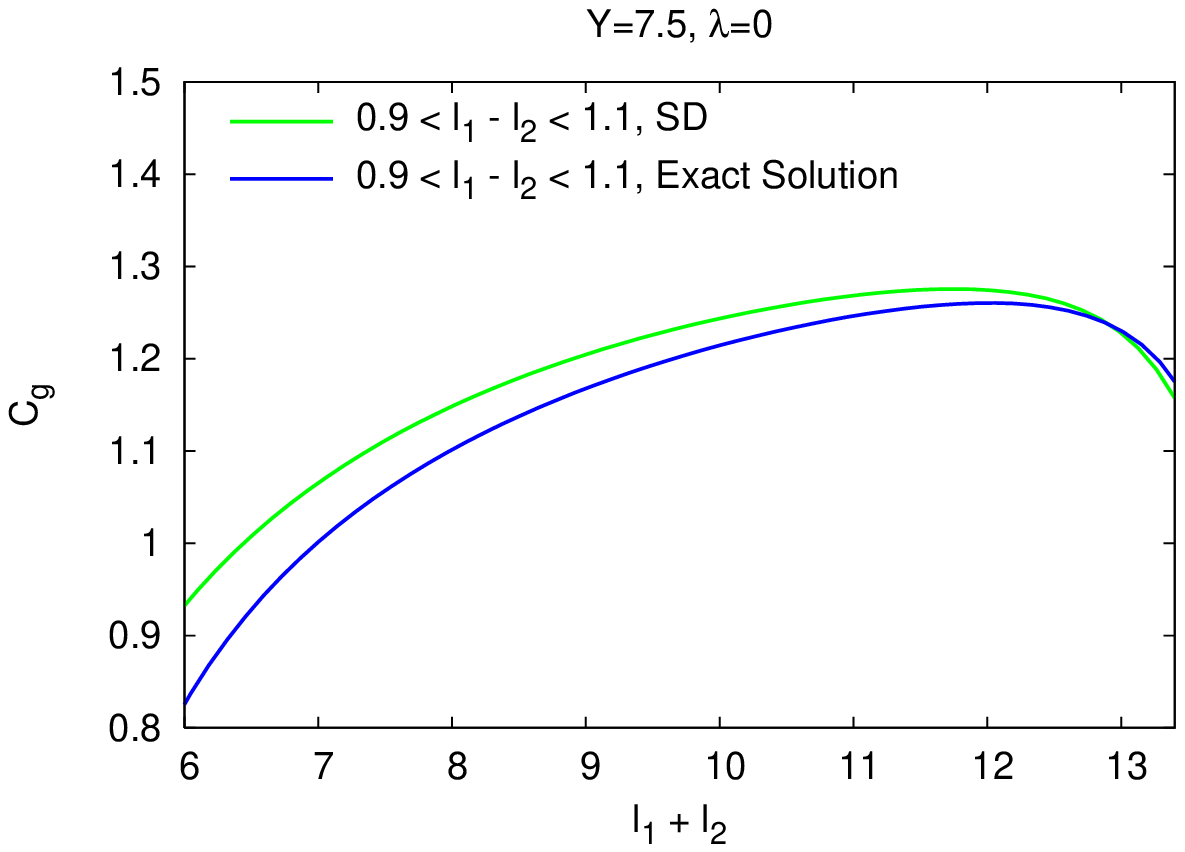, height=5truecm,width=0.48\tw}
\caption{\label{fig:MLLASTESpecter} Comparison between correlators
given by SD and in \cite{RPR2},  at $\lambda=0$.}
\end{center}
\end{figure}

\begin{figure}[h]
\begin{center}
\epsfig{file=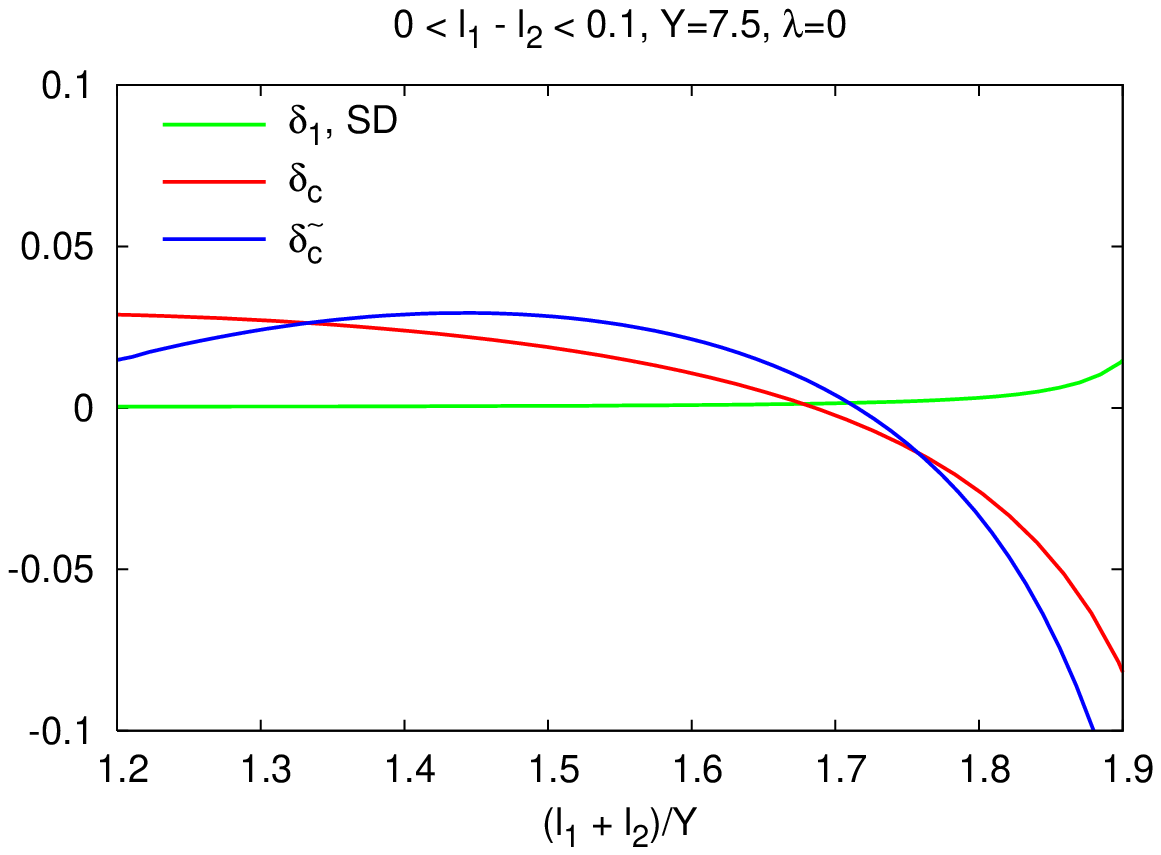, height=5truecm,width=0.48\tw}
\hfill
\epsfig{file=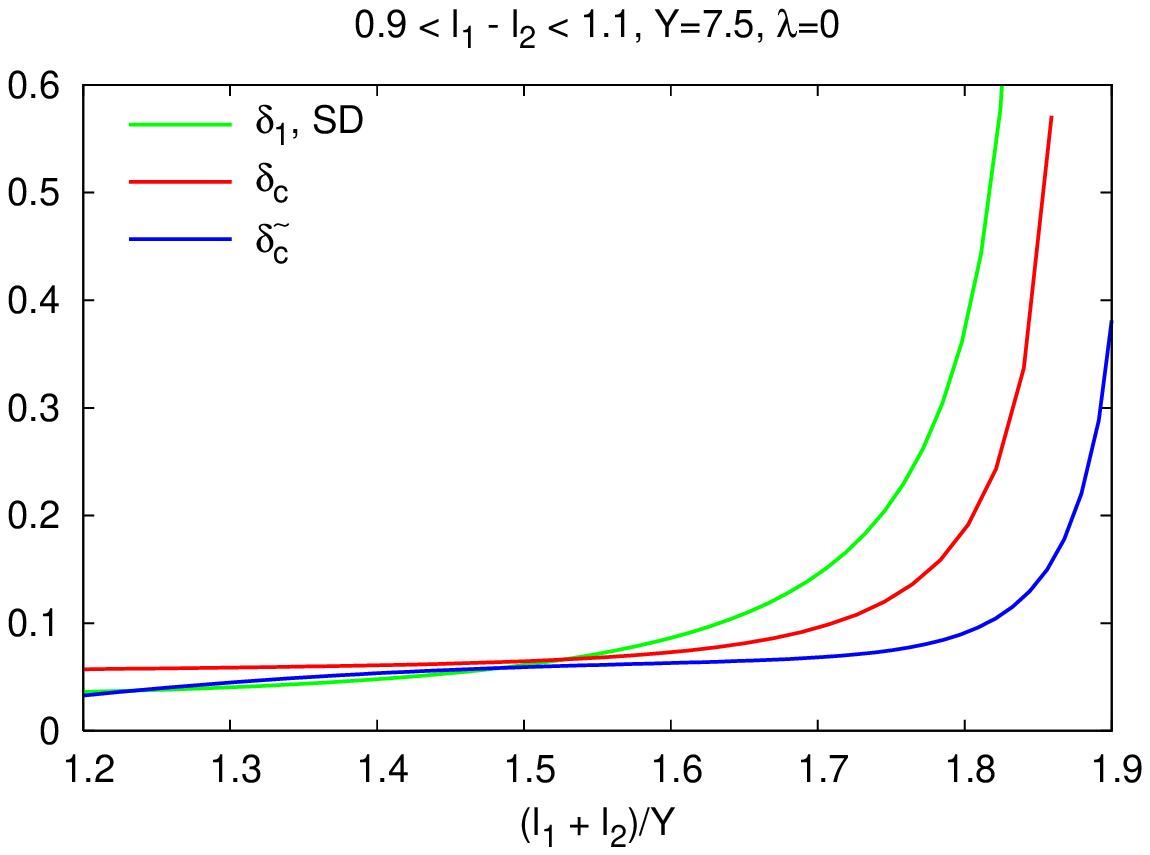, height=5truecm,width=0.48\tw}
\caption{\label{fig:deltas} Comparison between the SD $\delta_1$ and 
$\delta_c,\,\tilde\delta_c$ of \cite{RPR2} at $Y=7.5$, $\lambda=0$.}
\end{center}
\end{figure}

In Figs.\ref{fig:MLLASTESpecter} we compare the SD evaluation of the gluon correlator
with the exact solution of \cite{RPR2} at $\lambda=0$. The difference comes from sub-leading
corrections of order $\gamma_0^2$ that are not present in (\ref{eq:CGMLLAbis}). 
For example, $-\beta\gamma_0^2\approx-0.2$
at $Y=5.2$ occurring in the exact solution (69) of \cite{RPR2} 
is not negligible but is absent
in (\ref{eq:CGMLLAbis}) and (\ref{eq:rapMLLAbis}).
That is why, the SD MLLA curve lies slightly above 
the one of \cite{RPR2} at small $\ell_1+\ell_2$. The mismatch becomes smaller 
at $Y=7.5$, since $-\beta\gamma_0^2\approx-0.13$. However, when $\ell_1+\ell_2$
increases, the solution of \cite{RPR2} takes over, which can be explained by comparing
the behavior of the SD MLLA $\delta_1$ obtained in (\ref{eq:delta2}) and 
$\delta_c,\,\tilde\delta_c$ in \cite{RPR2}. Namely, while $\delta_1$ remains positive
and negligible for $\ell_1\approx\ell_2$, $\delta_c,\,\tilde\delta_c$ 
decrease and get negative when $\ell_1+\ell_2\to2Y$, see Fig.\ref{fig:deltas} (left),
which makes the correlations slightly bigger
in this region. As $|\ell_1-\ell_2|$ increases, $\delta_1$ is seen in
Fig.\ref{fig:deltas} (right) to play the same role as $\delta_c,\tilde\delta_c$ do
in the solution \cite{RPR2} and therefore, to decrease the correlation.
The agreement between both methods improves as the energy scale increases.
A similar behavior holds for the quark correlator.

In \cite{RPR2}, strong cancellations between the MLLA $\delta_1$ and the
NMLLA $\delta_2$  were seen to take place, giving very small
$\delta_c$ and $\tilde\delta_c$; this eased the
convergence of the iterative method but raised questions concerning the
relative size of MLLA and NMLLA corrections and the validity of the
perturbative expansion.
However, since $\delta_1$ is itself, there, entangled with {\em some} NMLLA
corrections, no definitive conclusions could be drawn.
The present work and Fig.~\ref{fig:deltas}, by showing that, below, 
$\delta_c$ and $\tilde \delta_c$ of \cite{RPR2} play the same role as
the {\em pure MLLA} $\delta_1$ which is now calculated, suggests (though it
is not a demonstration) that the perturbative series is safe.
It is indeed compatible with the following scheme: in \cite{RPR2},
the pure MLLA part of $\delta_1$ is the same as that in the present work;
the cancellations in \cite{RPR2} occur between NMLLA corrections
included in $\delta_1$ and $\delta_2$;  these are eventually of the same
order of magnitude as MLLA terms, but they are only parts of all
NMLLA corrections; this leaves the possibility that  the sum of all NMLLA
corrections to $\delta_1$ and all NMLLA terms of $\delta_2$ are separately
smaller than the pure MLLA terms of $\delta_1$, that is that strong
cancellations occur {\em between NMLLA corrections}, the ones
 included, because of the logic of the calculation, in \cite{RPR2}, and
those which were not be taken into account.

\subsection{Comparison with Fong-Webber and LEP-I data; how
$\boldsymbol{\lambda=0}$ is favored}
%%%%%%%%%%%%%%%%%%%%%%%%%%%%%%%%%%%%%%%%%%%%%%%%%%%%%%%%%%%%%%%%%%%%

Let us consider, at the $Z^0$ peak $Y=5.2$ ($E\Theta=91.2\,\text{GeV}\, \text{at}\, $LEP-I energy), the process $e^+e^-\to q\bar q$. As can be induced 
from Fig.\ref{fig:MLLASTESpecter},
the results obtained in the present work by the (approximate) SD method
are very close to the ones obtained in subsection 6.5 of \cite{RPR2}
by the exact solution of the evolution equations.
Accordingly, the same comparison as in \cite{RPR2} holds with respect to
both Fong \& Webber's results \cite{FW} and  OPAL data \cite{OPAL}.

It is also noticeable that, since, at $\lambda=0$, correlations already lye
above (present) experimental curves, and since an increase of
$\lambda$  tends to increase the predictions, the limiting spectrum stays the
best candidate to bring agreement with experiments.

%%%%%%%%%%%%%%%%%%%%%
\section{CONCLUSION}
\label{sec:CONCL}
%%%%%%%%%%%%%%%%%%%%%

Let us, in a few words, summarize the achievements,  but also the
limitations of the two methods that have been used respectively
in \cite{RPR2} (exact solution of MLLA evolution equations) and in the
present work (steepest descent approximate evaluation of their solutions).

Achievements are threefold:\newline
- in \cite{RPR2}, MLLA evolution equations for 2-particle
correlations have been deduced at small $x$ and at any $\lambda$; their
(iterative) solution can unfortunately only be expressed analytically at
the limit $\lambda \to 0$; 
\newline
- by the steepest descent method, which is an approximate method,
analytical expressions for the spectrum could instead be obtained
for $\lambda \ne 0$, which enabled to calculate the correlation at the same
level of generality; \newline
- one could move away from the peak of the inclusive distribution.

So doing, the limitations of the work of Fong \& Webber have vanished.
Their results have been recovered at the appropriate limits.

The two methods numerically agree remarkably well, despite an unavoidable
entanglement of MLLA + some NMLLA corrections in the first one.

The limitations are the following:\newline
- the uncontrollable increase of  $\alpha_s$ when one goes to smaller
and smaller transverse momenta: improvements in this directions mainly
concern the inclusion of non-perturbative contributions;\newline
- departure from the limiting spectrum: it cannot of course appear as a
limitation, but we have seen that increasing the value of $\lambda$, by
increasing the correlations, does not bring better agreement with present
data; it confirms thus, at present, that
the limiting spectrum is the best possibility;\newline
- the LPHD  hypothesis: it works surprisingly well for
inclusive distributions; only forthcoming data will assert
whether its validity decreases  when one studies less inclusive
processes (like correlations);\newline
- last, the limitation to small $x$: it is still quite drastic;
departing from this limit most probably lye in the art of numerical
calculations, which makes part of forthcoming projects.

Expectations rest on  experimental data, which are being collected at the
Tevatron, and which will be at LHC. The higher the energy, the safer
perturbative QCD is, and the better the agreement should be with our
predictions. The remaining disagreement (but much smaller than
Fong-Webber's) between predictions and LEP-1 results for 2-particle
correlations stands as an open
question concerning the validity of the LPHD hypothesis for these
observables which are not ``so'' inclusive as the distributions studied in
\cite{PerezMachet}. The eventual necessity to include NMLLA corrections can
only be decided when new data appear.

\vskip 1cm
\underline{\em Acknowledgments:}

It is a pleasure to thank Yuri Dokshitzer for enticing me towards the steepest
descent method and for showing me its efficiency with simple examples. I am
grateful to Bruno Machet for his help and advice and to Gavin Salam for helping
me in numerically inverting formula (\ref{eq:ratiomunu}).

%%%%%%%%%%%%%%%%%%%%%%%%%%%%%%%%%%%%%%%%%%%%%%%%%%%%%%%%%%%%%%%%%%%%%%%%%%%%%

\newpage
\appendix

%%%%%%%%%%%%%%%%%%%%%%%%%%%%%%%%%%%%%%%%%%%%%%%%%%%%%%%%%%%%%%%%%%%%%%%%%%%%%
\section{DOUBLE DERIVATIVES AND DETERMINANT}
\label{sec:SDdetails}
%%%%%%%%%%%%%%%%%%%%%%%%%%%%%%%%%%%%%%%%%%%%%%%%%%%%%%%%%%%%%%%%%%%%%%%%%%%%%

\subsection{Demonstration of eq.~(\ref{eq:determinant})}
\label{sec:DDDet}
%%%%%%%%%%%%%%%%%%%%%%%%%%%%%%%%%%%%%%%%%%%%%%%%%%%%

We conveniently rewrite (\ref{eq:deromega}) and (\ref{eq:dernu}) in the form

\begin{equation}\label{eq:derphiellbis}
\frac{\partial\phi}{\partial\omega}=\frac{2\omega-\nu}{\omega-\nu}\ell+
\frac{\nu}{\omega-\nu}-\frac{\phi}{\omega-\nu}-\lambda\frac{\nu+2s_0}{\omega-\nu}+
\frac1{\beta\omega(\omega-\nu)},
\end{equation}

\begin{equation}\label{eq:derphiybis}
\frac{\partial\phi}{\partial\nu}=\frac{\omega-2\nu}{\omega-\nu}y-
\frac{\omega}{\omega-\nu}+\frac{\phi}{\omega-\nu}+\lambda\frac{\omega+2s_0}{\omega-\nu}-
\frac1{\beta\nu(\omega-\nu)}.
\end{equation}
The Taylor expansion of (\ref{eq:phiexp}) in (\ref{eq:SpecDD}) reads
\begin{eqnarray}\nonumber
\phi(\omega,\nu,\ell,y)\!\!&\!\!\approx\!\!&\!\!\phi(\omega_0,\nu_0,\ell,y)+
\frac12\,\frac{\partial^2\phi}{\partial\omega^2}\,(\omega_0,\nu_0)(\omega-\omega_0)^2+
\frac12\,\frac{\partial^2\phi}{\partial\nu^2}\,(\omega_0,\nu_0)(\nu-\nu_0)^2\\\nonumber\\
&&+\frac{\partial^2\phi}{\partial\omega\partial\nu}
(\omega_0,\nu_0)\,(\omega-\omega_0)(\nu-\nu_0).
\end{eqnarray}

The expressions of the second derivatives follow directly from (\ref{eq:derphiellbis}) and (\ref{eq:derphiybis})
\begin{eqnarray}\nonumber
\frac{\partial^2\phi}{\partial\omega^2}=-\frac{\nu}{(\omega-\nu)^2}(\ell\!+\!y\!+\!\lambda)
+\frac{\phi}{(\omega-\nu)^2}-\frac{2\omega-\nu}{\beta\omega^2(\omega-\nu)^2}
+\frac4{\beta(\omega-\nu)^2(2s_0+\omega+\nu)},
\end{eqnarray}
\begin{eqnarray}\nonumber
\frac{\partial^2\phi}{\partial\nu^2}=-\frac{\omega}{(\omega-\nu)^2}(\ell\!+\!y\!+\!\lambda)
+\frac{\phi}{(\omega-\nu)^2}+\frac{\omega-2\nu}{\beta\nu^2(\omega-\nu)^2}
+\frac4{\beta(\omega-\nu)^2(2s_0+\omega+\nu)},
\end{eqnarray}
\begin{eqnarray}\nonumber
\frac{\partial^2\phi}{\partial\omega\partial\nu}=\frac{\omega}{(\omega-\nu)^2}(\ell\!+\!y\!+\!\lambda)
-\frac{\phi}{(\omega-\nu)^2}+\frac1{\beta\omega(\omega-\nu)^2}
-\frac4{\beta(\omega-\nu)^2(2s_0+\omega+\nu)}.
\end{eqnarray}

(\ref{eq:SpecDD}) and its solution can be written in the form

$$
G\simeq\iint d^2v\, e^{-\frac12 v^TAv}=\frac{2\pi}{\sqrt{Det\,A}}
$$
where 

$$
v=(\omega,\nu),\quad v^T=\left(\!\!
\begin{array}{c}
        \omega \cr
\cr
        \nu
 \end{array}
\!\!\right),\quad\begin{array}{c}
 DetA =\end{array} Det\left(
\begin{array}{c}
        \frac{\partial^2\phi}{\partial\omega^2}\qquad
 \frac{\partial^2\phi}{\partial\omega\partial\nu} \cr
\cr
        \frac{\partial^2\phi}{\partial\nu\partial\omega}\qquad
 \frac{\partial^2\phi}{\partial\nu^2}
 \end{array}
\right)=\frac{\partial^2\phi}{\partial\omega^2}\frac{\partial^2\phi}{\partial\nu^2}-
\left(\frac{\partial^2\phi}{\partial\omega\partial\nu}\right)^2. 
$$

An explicit calculation gives

$$
DetA=(\ell+y+\lambda)^2\left[\frac{\beta(\omega+\nu)\phi-4}{(\omega-\nu)^2}+\frac{4(\omega+\nu)}
{(\omega-\nu)^2(2s_0+\omega+\nu)}\right]
$$

which, by using (\ref{eq:muupsilon}) leads to (\ref{eq:determinant}).

\subsection{$\boldsymbol{DetA}$ (see eq.~(\ref{eq:determinant}))
around the maximum}
\label{subsec:DeTA}
%%%%%%%%%%%%%%%%%%%%%%%%%%%%%%%%%%%%%%%%%%%%%%%%%%%%%%%%%%%%%%%%%%%%%

This is an addendum to subsection \ref{subsection:SDeval} 

$\ell_{max}$ written in (\ref{eq:ellmaxmlla}) is close to the
DLA value $Y/2$ \cite{DLA}\cite{EvEq}\cite{KO}. We then have
$\mu\sim\upsilon\to0$ for $\ell\approx y\simeq Y/2$. In this limit,
(\ref{eq:ratiomunu})  and (\ref{eq:relmunu}) respectively translate into

\begin{equation}\label{eq:ellmu}
Y-2\ell\stackrel{\mu,\upsilon\to0}{\approx}\frac23\,
\frac{\left(Y+\lambda\right)^{3/2}-\lambda^{3/2}}
{\left(Y+\lambda\right)^{1/2}}\,\mu,\quad 
\upsilon\stackrel{\mu,\upsilon\to0}{\approx}
\sqrt{\frac{\lambda}{Y+\lambda}}\mu,
\end{equation}

while

\begin{equation}\label{eq:dermull}
\frac{\partial\mu}{\partial\ell}\simeq-3\frac{(Y+\lambda)^{1/2}}
{(Y+\lambda)^{3/2}-\lambda^{3/2}}
\end{equation}

should be used to get (\ref{eq:gaussian}). An explicit calculation gives

$$
\lim_{\mu,\upsilon\rightarrow0}
\sqrt{\frac{\beta^{1/2}(Y+\lambda)^{3/2}}{\pi DetA(\mu,\upsilon)}}\!=\!
\left(\frac{3}{\pi\sqrt{\beta}\left[(Y+\lambda)^{3/2}
\!-\!\lambda^{3/2}\right]}\right)^{1/2},
$$

where

\begin{eqnarray}\nonumber
DetA\!&\!\stackrel{\mu,\upsilon\to0}\approx
\!&\!\beta(Y\!+\!\lambda)^3\frac{\left(\mu\!-\!\upsilon\right)
\left(1\!+\!\frac12
\mu^2\right)\left(1\!+\!\frac12\upsilon^2\right)\!+\!(1\!+\!\frac12
\mu^2)\left(\upsilon\!+\!\frac16\upsilon^3\right)\!-\!\left(\mu\!+\!\frac16\mu^3\right)
\left(1\!+\!\frac12\upsilon^2\right)}{\mu^3}\\\notag\\
\!&\!\simeq\!&\!\frac13\beta(Y\!+\!\lambda)^3\left(1\!-\!\frac{\upsilon^3}{\mu^3}\right)=
\frac13\beta(Y+\lambda)^3\left[1-\left(\frac{\lambda}{Y+\lambda}\right)^{3/2}\right].
\end{eqnarray}

\subsection{The functions $\boldsymbol{L(\mu,\upsilon)}$, 
$\boldsymbol{K(\mu,\upsilon)}$ in eq.~(\ref{eq:LK})}
\label{subsec:LK}
%%%%%%%%%%%%%%%%%%%%%%%%%%%%%%%%%%%%%%%%%%%%%%%%%%%%%%%%%

An explicit calculation gives
\begin{equation}\label{eq:LL}
L(\mu,\upsilon)=\frac32\frac{\cosh\mu}{\sinh\mu}-
\frac12\frac{(\mu-\upsilon)\cosh\upsilon\sinh\mu+\sinh\upsilon\sinh\mu}
{(\mu-\upsilon)\cosh\mu\cosh\upsilon+\cosh\mu\sinh\upsilon-\sinh\mu\cosh\upsilon},
\end{equation}
and
\begin{equation}\label{eq:KK}
K(\mu,\upsilon)=-\frac12\sinh\upsilon\frac{(\mu-\upsilon)\cosh\mu-\sinh\mu}
{(\mu-\upsilon)\cosh\mu\cosh\upsilon+
\cosh\mu\sinh\upsilon-\sinh\mu\cosh\upsilon}.
\end{equation}

\subsection{A consistency check}
\label{subsec:check}
%%%%%%%%%%%%%%%%%%%%%%%%%%%%%%%%

Let us verify that the evolution equation (\ref{eq:solg}) is satisfied
by (\ref{eq:SpecNormMLLA}) within the MLLA accuracy.
Differentiating (\ref{eq:solg}) with respect to $\ell$, $y$
yields the equivalent differential equation

$$
G_{\ell y} = \gamma_0^2 \left(G - a G_\ell\right) \!+\! {\cal {O}}{\big(\gamma_0^4G\big)}
$$

that can be rewritten in the form

\begin{equation}\label{eq:MLLApsi}
 \psi_\ell\psi_y +\psi_{\ell y} \>=\>
 \gamma_0^2\left(1-a\psi_\ell\right)  \>+\> {\cal {O}}\big({\gamma_0^4}\big);
\end{equation}

we have neglected next-to-MLLA corrections ${\cal O}(\gamma_0^4)$
(of relative order $\gamma_0^2$) coming from differentiating the coupling
$\gamma_0^2$ in the sub-leading (``hard correction'') term $\propto a$.

We have to make sure that (\ref{eq:MLLApsi}) holds including the terms
${\cal O}(\gamma_0^3)$. In the sub-leading terms we can set $\psi\to \varphi$
(see (\ref{eq:SPbis})):

\begin{equation}
  (\varphi_\ell + \delta\psi_\ell)(\varphi_y + \delta\psi_y) + \varphi_{\ell
  y} = \gamma_0^2(1-a\varphi_\ell). 
\end{equation}

Isolating correction terms and casting them all on the l.h.s. of the
equation we get

\begin{equation}\label{eq:collect}
  a\gamma_0^2\varphi_\ell 
+ \left[\,\varphi_\ell \delta\psi_y + \varphi_y\delta\psi_\ell\,\right] + \varphi_{\ell
  y} \>=\>  \gamma_0^2 -\varphi_\ell\varphi_y.
\end{equation}

By the definition (\ref{eq:SPbis}) of the saddle point
we conclude that the r.h.s. of (\ref{eq:collect}) is zero such that we have

\begin{equation}\label{eq:hastobe}
   \omega_0 a\gamma_0^2 + \left[\,\omega_0 \delta\psi_y +
     \nu_0\delta\psi_\ell\,\right] + \frac{d\omega_0}{dy} \>=\> 0\,,
\end{equation}

that is,

\begin{equation}\label{eq:omega0}
   \omega_0 \left(a\gamma_0^2 + \delta\psi_y\right) +
     \nu_0\delta\psi_\ell + \frac{d\omega_0}{dy} \>=\> 0\,.
\end{equation}

First, we select the terms $\propto a$:

\begin{eqnarray*}
&&a\gamma_0^3\left[-\frac12\widetilde{Q}-\frac12\tanh\upsilon\, e^{\mu}
+\frac12\tanh\upsilon\coth\mu\, e^{\mu} + \frac12\tanh\upsilon\coth\mu \,
\widetilde{Q}\right.\\
&&\left.+\frac12\widetilde{Q}-\frac12\tanh\upsilon\,e^{-\mu}-\frac12
\tanh\upsilon\coth\mu\, e^{-\mu}-\frac12\tanh\upsilon\coth\mu \,
\widetilde{Q}\right]\\
&&=a\gamma_0^3\left[-\tanh\upsilon\cosh\mu+\tanh\upsilon\coth\mu\sinh\mu\right]
\equiv0.
\end{eqnarray*}

From (\ref{eq:muupsilon}) one deduces

$$
\frac{d\omega_0}{dy}=\frac12\beta\gamma_0^3\widetilde{Q},
$$

that is inserted in (\ref{eq:omega0}) such that,
for terms $\propto\beta$, we have

\begin{eqnarray*}
&&-\beta\gamma_0^3\left[\frac12e^{\mu}+\frac12\tanh\upsilon
\Big(1\!+\!K\Big)e^{\mu}-\frac12C\,e^{\mu}-\frac12C\widetilde{Q}+
\frac12e^{-\mu}+\frac12\tanh\upsilon
\Big(1\!+\!K\Big)e^{-\mu}\right.\\
&&\left.+\frac12C\,e^{-\mu}+\frac12C\widetilde{Q}\right]
=-\beta\gamma_0^3\left[\cosh\mu+\tanh\upsilon\cosh\mu\Big(1\!+\!K\Big)
-C\sinh\mu-\frac12\widetilde{Q}\right],
\end{eqnarray*}

which gives

\begin{eqnarray*}
&&-\beta\gamma_0^3\left[\cosh\mu-\sinh\mu\,L-\frac12\widetilde{Q} \right].
\end{eqnarray*}

 Constructing (see (\ref{eq:tildeQ}) and appendix \ref{subsec:LK})

\begin{eqnarray*}
\widetilde{Q}(\mu,\upsilon)-2\cosh\mu\!\!&\!\!=\!\!&\!\!
-3\cosh\mu+\sinh\mu\frac{(\mu-\upsilon)\cosh\upsilon\sinh\mu+\sinh\upsilon\sinh\mu}
{(\mu-\upsilon)\cosh\mu\cosh\upsilon+\cosh\mu\sinh\upsilon-\sinh\mu\cosh\upsilon}\\
\!\!&\!\!=\!\!&\!\!-2\sinh\mu L(\mu,\upsilon)
\end{eqnarray*}

we have

$$
-\beta\gamma_0^3\left[\cosh\mu-\sinh\mu\,L-\frac12\widetilde{Q}\right]\equiv0.
$$

%%%%%%%%%%%%%%%%%%%%%%%%%%%%%%%%%%%%%%%%%%%%%%%%%%%%%%%%%%%%%%%%%%%%%%%
\section{ANALYTICAL EXPRESSION  OF $\boldsymbol{\Delta'(\mu_1,\mu_2)}$
OBTAINED FROM  EQ.~(\ref{eq:Deltaprime})}
\label{subsec:Deltaprime}
%%%%%%%%%%%%%%%%%%%%%%%%%%%%%%%%%%%%%%%%%%%%%%%%%%%%%%%%%%%%%%%%%%%%%%%

Replacing (\ref{eq:derpsi'l})(\ref{eq:derpsi'y}) in
(\ref{eq:Deltaprime}) and neglecting terms of relative order ${\cal O}(\gamma_0^3)$
which are beyond the MLLA accuracy, we obtain

\begin{eqnarray}\label{eq:dDelta}
\Delta'\!\!&\!\! = \!\!&\!\!
\frac{ e^{-\mu_1}\delta\psi_{2,\ell} + e^{-\mu_2}\delta\psi_{1,\ell} 
       + e^{\mu_1}\delta\psi_{2,y} + e^{\mu_2}\delta\psi_{1,y}} {\gamma_0}\cr\cr
\!\!&\!\!=\!\!&\!\!-a\gamma_0\left[e^{\mu_1}+e^{\mu_2}-\sinh(\mu_1-\mu_2)(\widetilde{Q}_1-
\widetilde{Q}_2)+\cosh\mu_1\tanh\upsilon_2+\cosh\mu_2\tanh\upsilon_1\right.\cr\cr
&&\hskip 1cm\left.-\sinh\mu_1\tanh\upsilon_2\coth\mu_2-
\sinh\mu_2\tanh\upsilon_1\coth\mu_1\right.\cr\cr
&&\hskip 1cm\left.+\sinh(\mu_1-\mu_2)\Big(\tanh\upsilon_1
\coth\mu_1\widetilde{Q}_1-\tanh\upsilon_2
\coth\mu_2\widetilde{Q}_2\Big)\right]\cr\cr
&&-\beta\gamma_0\left[\Big(\cosh\mu_1-\sinh\mu_1C_2\Big)+\Big(\cosh\mu_2-\sinh\mu_2C_1\Big)
+\sinh(\mu_1-\mu_2)(C_1\widetilde{Q}_1-C_2\widetilde{Q}_2)\right.\cr\cr
&&\hskip 1cm\left.+\cosh\mu_1\tanh\upsilon_2\Big(1+K_2\Big)+\cosh\mu_2\tanh\upsilon_1
\Big(1+K_1\Big)\right].
\end{eqnarray}

%%%%%%%%%%%%%%%%%%%%%%%%%%%%%%%%%%%%%%%%%%%%%%%%%%%%%%%%%%%%%%%%%%%%%%%%%%%%%%
\null

\vskip 2cm

\listoffigures

%%%%%%%%%%%%%%%%%%%%%%%%%%%%%%%%%%%%%%%%%%%%%%%%%%%%%%%%%%%%%%%%%%%%%%%%%%%%%%

%%%%%%%%%%%%%%%%%%%%%%%%%%%%%%%%%%%%%%%%%%%%%%%%%%%%%%%%%%%%%%%%%%%%%%%%%%%


\newpage

\begin{thebibliography}{50}
%
\bibitem{RPR2}
R. Perez-Ramos:
%``Two particle correlations inside one jet at ``Modified Leading
%logarithmic Approximation'' of Quantum Chromodynamics;
%I: Exact solution of the evolution equations at small $x$,
JHEP 06 (2006) 019, hep-ph/0605083, and references therein.

\bibitem{DKT}
Yu.L. Dokshitzer, V.A. Khoze and S.I. Troyan: Int. J. Mod.
Phys. {\bf A7} (1992) 1875.

\bibitem{FW} C.P. Fong and B.R. Webber: Phys. Lett. {\bf B 241} (1990) 255.

\bibitem{DKTM}
Yu.L. Dokshitzer, V.A. Khoze, S.I. Troyan and A.H. Mueller: Rev. Mod.
Phys. {\bf 60} (1988) 373-388.

\bibitem{PerezMachet}
R. Perez-Ramos \& B. Machet:
%``MLLA inclusive hadronic distributions inside one jet at high energy
%colliders'',
JHEP 04 (2006) 043, hep-ph/0512236.

\bibitem{OPAL} OPAL Collab.: Phys. Lett. {\bf B 287} (1992) 401.

\bibitem{DLA} Yu.L. Dokshitzer, V.S. Fadin and V.A. Khoze: Z. Phys
  {\bf C18} (1983) 37.

\bibitem{EvEq} 
Yu.L. Dokshitzer, V.A. Khoze, A.H. Mueller and S.I. Troyan:
{\em ``Basics of Perturbative QCD''}, Editions Fronti\`eres, Paris, 1991.

\bibitem{KO} V.A. Khoze and W. Ochs: Int. J. Mod. Phys. {\bf A12}
  (1997) 2949.

\bibitem{FW1} C.P. Fong and B.R. Webber: Phys. Lett. {\bf B 229} (1989) 289.


\end{thebibliography}
\end{document}